\documentclass[12pt]{article}
\usepackage{amsmath,mathrsfs,amsthm,amsfonts}
\usepackage{graphicx}
\usepackage{enumerate}
\usepackage{natbib}
\usepackage{url} 
\usepackage{xcolor}
\usepackage{cleveref}

\newcommand{\blind}{1}

\def\mC{{\mathscr C}}
\def\gamma{\beta \!\!\! / }
\def\anova{\hat\theta_{{}_{\rm AN}}}

\def\ancova{\hat\theta_{{}_{\rm ANC}}}
\def\hancova{\hat\theta_{{}_{\rm ANHC}}}
\def\shancova{\hat\theta_{{}_{\rm ANHC}}}

\def\var{{\rm var}}

\crefname{equation}{}{}

\newtheorem{theorem}{Theorem}
\crefname{theorem}{Theorem}{Theorems}
\newtheorem{lemma}{Lemma}
\crefname{lemma}{Lemma}{Lemmas}

\crefname{requirement}{Consideration}{Considerations}
\newtheorem{corollary}{Corollary}

\begin{document}

\def\spacingset#1{\renewcommand{\baselinestretch}%
  {#1}\small\normalsize} \spacingset{1}


\if1\blind
{
  \begin{center}
    {\Large\bf  Toward Better Practice of Covariate Adjustment \vspace{2mm} \\ in Analyzing Randomized Clinical Trials} \\ \bigskip \bigskip
    {\large  Ting Ye\footnote{Department of Biostatistics, University of Washington.}, Jun Shao\footnote{School of Statistics, East China Normal University; Department of Statistics, University of Wisconsin-Madison.},
	Yanyao Yi\footnote{Global Statistical Sciences, Eli Lilly and Company. }, and Qingyuan Zhao\footnote{Department of Pure Mathematics and
		Mathematical Statistics, University of Cambridge.\\ Corresponding to Dr. Jun Shao. Email: {\tt shao@stat.wisc.edu}.}
}
  \end{center}
} \fi

\if0\blind
{
  \bigskip
  \bigskip
  \bigskip
  \begin{center}
    {\LARGE\bf Toward better practice of Covariate Adjustment in Analyzing Randomized Clinical Trials}
  \end{center}
  \medskip
} \fi

\bigskip
\begin{abstract}
 In randomized clinical trials, adjustments for baseline covariates at both
design and analysis stages are highly encouraged by
regulatory agencies. A recent trend is to use a model-assisted
approach for covariate adjustment to gain credibility and efficiency while producing
asymptotically valid inference even when the model is incorrect. In this
article we present three considerations for better practice when model-assisted
inference is applied to adjust for covariates under simple or covariate-adaptive randomized trials: (1)  guaranteed efficiency gain: a model-assisted method should often gain but never hurt efficiency; (2) 
wide applicability: a valid procedure should be 
applicable, and preferably 
universally applicable, to all commonly used randomization schemes; (3) robust standard error:  variance estimation should be  robust to model misspecification and heteroscedasticity. 
	To achieve these, we recommend a model-assisted estimator under 
 an analysis of heterogeneous 
	covariance working model including all covariates utilized in randomization. 	
	Our conclusions are based on  an asymptotic theory 
that provides a clear picture of how covariate-adaptive
randomization and regression adjustment alter statistical
efficiency.  
Our theory is more general than the existing ones  
in terms of studying arbitrary  functions of response means (including linear contrasts, ratios, and odds ratios), multiple arms,  guaranteed efficiency gain, optimality, and universal applicability.

\end{abstract}

\noindent%
{\it Keywords:}  Analysis of covariance;
Covariate-adaptive randomization;
Efficiency; 
Heteroscedasticity; Model-assisted; Multiple treatment arms; Treatment-by-covariate interaction.
\vfill

\newpage
\spacingset{1.5} 
\section{Introduction}
\label{sec:intro}

Consider a clinical trial with patients randomized into one and only one of multiple treatment arms according to fixed assignment proportions. Each patient has multiple potential responses, one for each treatment, but 
only one response is observed
 depending on the assigned treatment. 
Based on data collected from the trial, we would like to make statistical inference on  treatment effects defined as functions of the response means (e.g., linear contrasts, ratios, or odds ratios).  
These unconditional treatment effects are discussed in a recent Food and Drug Administration draft guidance  \citep{fda:2019aa}.




In clinical trials, we typically observe some baseline covariates for each
patient, which  are measured prior to treatment assignments and, hence, are not affected by
the treatment.  As emphasized in regulatory agency guidelines, baseline covariates are encouraged to be utilized in the following  two
ways. 
(i) In the design stage, covariate-adaptive randomization can be used to
enforce the balance of treatment assignments across levels of discrete baseline prognostic factors, such as
institution, disease stage, prior treatment, gender, and age
group. ``Balance of treatment groups with respect to one or more
specific prognostic covariates can enhance the credibility of the
results of the trial'' \citep[European Medicines Agency]{ema:2015aa}. (ii)  In the analysis stage, baseline
covariates can be used to gain efficiency. 
``Incorporating prognostic baseline factors in the primary statistical analysis of clinical trial data can result in a more efficient use of data to demonstrate and quantify the effects of treatment with minimal impact on bias or the Type I error rate''
\citep{fda:2019aa}. More specifically, the investigator is advised to
 ``identify those covariates and factors 
expected to have an important influence on the primary variable(s)'' and to specify
``how to account for them in the analysis in order to
improve precision and to compensate for any lack of balance between
groups'' \citep{ICHE9}. 

 For efficiency gain, one may apply a \emph{model-based} approach using  a model  between the potential responses and covariates.  However, the validity of a model-based approach requires a correct model specification, i.e.,  
 a possibly strong assumption. 
 As emphasized in \cite{fda:2019aa}, 
 a method used for covariate adjustment ``should provide valid inference under approximately the same minimal statistical assumptions that would be needed for unadjusted estimation in a randomized trial''. 
Consequently, 
\emph{model-assisted}
approaches,
which gain efficiency  through a working model between responses and covariates and still produce
asymptotically valid inference even when the working
model is misspecified,  have become considerably more popular.


\subsection{Considerations about covariate adjustment}

For better practice of covariate adjustment via model-assisted approaches, we present the following three considerations. 

\noindent
{\bf 1. Guaranteed efficiency gain}.
{\em 	The working model should be chosen so that the resulting model-assisted estimator
	often gains but never loses efficiency when compared to a benchmark
	estimator that does not adjust for any covariate.}

This consideration is  important for model-assisted inference because covariate adjustment based on a misspecified working model does not necessarily lead to efficiency gain over the benchmark.
One example is  the customary analysis of covariance (ANCOVA) whose working model does not include
treatment-by-covariate interaction terms, which we refer to as the \emph{homogeneous} working model (\S 2.3). These interaction terms are often ignored or even discouraged in practice because of two perceptions: (i) even if the homogeneous 
working model is misspecified, ANCOVA still provides valid inference
as it is model-assisted; (ii) a model without interaction terms have
fewer coefficients to estimate and may have better finite sample
properties. 
These perceptions are correct but only provide a partial picture. 
When the treatment effect is indeed heterogeneous,  the 
ANCOVA  estimator 
using the homogeneous working model  may be even less efficient than
the benchmark analysis of variance (ANOVA) estimator that uses no model assistance at all \citep{Freedman:2008aa, Lin:2013aa}. 
This has led to confusion about how covariate adjustment should be implemented, which can be
seen from  conflicting recommendations by regulatory agencies:
``The primary model should not include treatment
by covariate interactions.''  \citep{ema:2015aa}; 
``The linear models may include treatment by covariate interaction terms.'' \citep{fda:2019aa}. 

{Is there a model-assisted method that achieves guaranteed efficiency gain?
An affirmative answer is provided in \S1.2. }

\noindent
{\bf 2. Wide applicability}.
{\em	The model-assisted inference procedure should be applicable to all commonly used randomization schemes}. 

Covariate-adaptive randomization has been widely used in modern clinical trials to balance treatments across important prognostic factors.  
According to a recent review of nearly 300 clinical trials published in 2009
and 2014, 237 of them used covariate-adaptive randomization \citep{Ciolino:2019aa}. The three most popular
covariate-adaptive randomization schemes are 
the stratified permuted block  \citep{Zelen:1974aa},  the stratified 
biased coin  \citep{Shao:2010aa,Kuznetsova:2017aa}, 	and 
Pocock-Simon's minimization \citep{Taves:1974aa, Pocock:1975aa,Han:2009aa}.  Unlike simple randomization, covariate-adaptive randomization generates a dependent sequence of treatment assignments. As recognized by regulatory agencies \citep{ema:2015aa,fda:2019aa}, 
conventional inference procedures developed under simple randomization are not necessarily valid under covariate-adaptive randomization. 
Thus, the second consideration is whether the model-assisted inference procedure is applicable to all commonly used randomization schemes.


\noindent
{\bf 3. Robust standard error}. 
{\em 	The model-assisted inference should use  standard errors  robust against model misspecification and heteroscedasticity.}

The use of robust standard error is a
 crucial step for valid model-assisted inference \citep{fda:2019aa}. Although the asymptotic theory for heteroscedasticity-robust standard errors
was developed decades ago
\citep{Huber:1967aa,White:1980aa} and has been widely used in
econometrics, its usage in clinical trials is scarce.

\subsection{Our contributions}

Given how frequently covariate adjustment is being used in practice, it may come as a surprise that there has been no comprehensive guideline yet. 
In our opinion, this is because most existing papers consider some aspects but not a full picture regarding the three considerations described in \S1.1. 
For example, most existing results are for linear contrasts of response means for two arms  \citep[among
others]{Yang:2001aa, Tsiatis:2008aa, Shao:2010aa, Lin:2013aa, Shao:2013aa,Ma:2015aa, Bugni:2017aa,Ye:2018aa, Wang:2019aa, wang2020modelrobust,Liu:2020aa,Ma:2020aa,ma2020regression};
many of them  are applicable to only simple randomization or a limited class of randomization schemes   with  well-understood properties;
\cite{Bugni:2019aa} and \cite{Ye:2020ab} consider multiple arms but still focus on linear contrasts and do not fully address the guaranteed efficiency gain or optimality;  not enough insights are provided to convince practitioners to apply model-assisted inference.


 We establish a comprehensive theory to 
 provide a clear picture 
of how covariate-adaptive
randomization and regression adjustment alter  statistical
efficiency, which resolves some confusion about covariate adjustment  
and facilitates its better practice
with easy-to-implement recommendation for practitioners. 
Our theory is more general than the existing ones  
in terms of studying arbitrary functions of response means (including linear contrasts, ratios and odds ratios), multiple arms, guaranteed efficiency gain, optimality, and universal applicability.


	
	Our theory shows that a  \emph{heterogeneous} working model for ANCOVA that includes all treatment-by-covariate interaction terms should be favored because it achieves both guaranteed efficiency gain and wide applicability when all covariates utilized in covariate-adaptive randomization are included in the working model. 
To distinguish from the customary ANCOVA that uses a homogeneous working model, we term the ANCOVA using a heterogeneous working model as ANalysis of HEterogeneous COVAriance (ANHECOVA). Note that ANHECOVA is not a new proposal and has a long history in the literature with a recent resurgence of attention \citep[among
	others]{Cassel:1976aa, Yang:2001aa, Tsiatis:2008aa, Lin:2013aa, Wang:2019aa, Liu:2020aa,Li:2020aa}, but our recommendation of ANHECOVA is from a more comprehensive perspective. Specifically, 
	 in \S\ref{sec: inference SR}-\S\ref{sec: inference CAR},  we show that 
	 under mild and transparent assumptions, the recommended ANHECOVA estimator of the response mean vector is consistent, asymptotically normal, and asymptotically more efficient than the benchmark ANOVA or ANCOVA estimator;	in fact, the ANHECOVA estimator is asymptotically the most efficient estimator within a
	 class of linearly-adjusted estimators. Special cases of this result have been discussed in the literature, but our development is for a much more general setting that considers multiple treatment arms,  joint estimation of response means, and under all commonly used covariate-adaptive randomization schemes.  In \S3.1 we offer  explanations of why the heterogeneous working model is generally preferable over the homogeneous working model.

Besides guaranteed efficiency gain and wide applicability, our asymptotic theory in  \S3.2-3.3 shows that the recommended ANHECOVA procedure also enjoys a \emph{universality} property, i.e.,   the same inference procedure can be universally applied to all commonly used randomization schemes including Pocock-Simon's minimization whose asymptotic property is still not well understood.  
This is because the asymptotic variance of the ANHECOVA estimator is invariant to the
randomization scheme, as long as the randomization scheme satisfies a very mild condition (C2) stated in \S2.2. 
The universality property is desirable for practitioners as 
they do not need to derive a tailored standard error formula 
for each randomization scheme.


The standard  heteroscedasticity-robust standard error formulas 
do not directly apply to model-assisted inference for clinical trials because they do not take into account covariate centering prior to model fitting. 
In \S 3.4, we develop a robust standard error formula that can be used with the ANHECOVA estimator. 

	Finally, our investigation offers new insights on when ANCOVA as a model-assisted inference approach can achieve guaranteed efficiency gain over the benchmark  ANOVA. For example, under simple randomization
	with two treatment arms, \citet{Lin:2013aa}
	showed that ANCOVA  has this desirable property  if inference focuses on a linear contrast  and the
	treatment allocation is balanced. However, our theory shows  that this
	does not extend to trials with more than
	two  arms or inference on nonlinear functions  of response means (such as ratios or odds ratios), and is thus a peculiar property for ANCOVA. In addition, ANCOVA does not have wide  applicability  because the asymptotic normality of the ANCOVA estimator requires an additional condition (C3) on randomization, which is not satisfied by the popular Pocock-Simon's minimization method.
	Even when ANCOVA is applicable to a particular randomization scheme, it does not have universality because its asymptotic variance varies with the randomization scheme \citep{Bugni:2017aa}.

%

After introducing the notation, basic assumptions, and working models in
\S2, we present the methodology and theory in \S3.  Some numerical
results are given in \S4. The paper is concluded with recommendations 
and discussions for clinical trial practice in
\S5. Technical proofs can be found in the supplementary material.
\vspace{-3mm}

\section{Trial Design and Working Models}
\label{sec: design}
\vspace{-2mm} 

\subsection{Sample}
\vspace{-2mm} 

In a clinical trial with $ k $ treatment arms, let $Y^{(t)}$ represent the
potential  (discrete or continuous) response under treatment $t$, 
$t = 1, \dotsc, k$, $ \theta $
be the $ k $-dimensional vector whose $ t $th component is $ \theta_t= E(Y^{(t)}) $, the unknown potential response mean under treatment $ t $, where $ E $ denotes the population expectation. We are interested in given functions of $ \theta $, such as a linear contrast $ \theta_t- \theta_s $, a ratio $ \theta_t/\theta_s $, or an odds ratio $ \{\theta_t/(1-\theta_t)\} / \{ \theta_s/(1-\theta_s)\}$ between two treatment arms $ t $ and $ s $.  We  use
$Z$ to denote the vector of discrete baseline covariates used in
covariate-adaptive randomization and $X$ to denote the vector of baseline
covariates used in model-assisted inference. The vectors $Z$ and $X$
are allowed to share the same entries.

Suppose that a random sample of $n$ patients is obtained from the
population under investigation. 
For the $i$th patient,
let $Y^{(1)}_i,..., Y_i^{(k)}$, $Z_i$, and $ X_i $ be the realizations of
$Y^{(1)},..., Y^{(k)}$, $Z$, and $X$, respectively. We impose the following
mild condition. \vspace{-1mm} 
\begin{itemize}
	\item[(C1)] $(Y_i^{(1)}, \dotsc, Y_i^{(k)}, Z_i, X_i)$,
	$i=1,\dotsc,n$, are
	independent and identically distributed with finite second order 
	moments. The distribution of baseline covariates is not affected by treatment and the covariance matrix  $\Sigma_X = {\rm var}(X_i)$   is 
		positive definite. \vspace{-1mm} 
\end{itemize}
Notice that neither a model between the potential responses and baseline covariates 
nor a distributional assumption on 
potential responses is assumed. 
\vspace{-1mm} 

\subsection{Treatment assignments}
\vspace{-2mm} 

Let $\pi_1,\dotsc,\pi_k$ be the pre-specified treatment assignment
proportions,  $0 < \pi_t < 1$, and $\sum_{t=1}^k \pi_t = 1$. Let 
  $A_i  $  be the $k$-dimensional treatment indicator vector
that equals $a_t$ if patient $i$ receives treatment $t$, where $a_t$
denotes the $k$-dimensional vector whose $t$th component is 1 and
other components are 0.
For patient $i$, only one treatment is assigned according to $A_i$ after baseline covariates
$Z_i$ and $X_i$ are observed. The
observed response is $Y_i = Y_i^{(t)}$ if and only if $ A_i=a_t $. 
Once the treatments are assigned and the
responses are recorded, the statistical inference is based on the
observed $(Y_i, Z_i, X_i, A_i)$ for $i=1,...,n$.

The simple randomization scheme assigns patients to  treatments completely at random, under which  $A_i$'s are independent of  $ (Y_i^{(1)},..., Y_i^{(k)}, X_i)$'s and are independent and identically distributed with $P(A_i =a_t)= \pi_t$, $t=1,...,k$.   It  does not make use of covariates and, hence,
may yield sample sizes that substantially deviate from the target assignment proportions across  levels of the prognostic factors. 

To improve the credibility of the trial, it is often desirable to
enforce the targeted treatment assignment proportions across levels of
 $Z$ by using covariate-adaptive randomization. 
As introduced in Section \ref{sec:intro}, the three most popular
 covariate-adaptive randomization schemes   are 
 the stratified permuted block and stratified 
 biased coin, both of which 
 use all joint levels of  $Z$ as strata, 	and 
 Pocock-Simon's minimization, 
 which aims to enforce treatment assignment proportions across marginal levels of  $ Z$.

All  these covariate-adaptive randomization schemes, as well as the simple randomization, 
satisfy the following mild condition \citep{Baldi-Antognini:2015aa}.\vspace{-2mm} 
\begin{itemize}
	\item[(C2)] The discrete covariate $Z$ used in randomization has finitely many joint levels in $\mathcal{Z}$ and satisfies (i) given $\{Z_i,i=1,...,n\}$,
		$\{A_i,i=1,...,n\}$ is conditionally independent of
		$\{(Y_i^{(1)},..., Y_i^{(k)}, X_i), i=1,...,n\}$;  (ii) as $n \to
	\infty$,  $n_t (z) /n(z) \to  \pi_t $  almost surely, where $n(z)$
	is the number of patients with $Z=z$ and $n_t(z)$ is the number of
	patients with $Z=z$ and treatment $t$, $z\in \mathcal{Z}$,
	$t=1,...,k$.\vspace{-2mm} 
\end{itemize}

\subsection{Working models}

\label{sec:working-models}

\vspace{-2mm} 

The ANOVA  considered as benchmark throughout this paper does not model how the potential responses
$Y_i^{(1)},...,Y_i^{(k)}$ depend on the baseline covariate vector $X_i$.  It is based \vspace{-2mm} on
\begin{equation}
  \label{anova}
  E(Y_i \mid A_i) = \vartheta^T A_i, \vspace{-2mm}
\end{equation}
where $\vartheta$ is a $k$-dimensional unknown  vector and 
$c^T$ denotes the row vector that is the transpose
of a column vector $c$. By Lemma 2 in the supplementary material, $\vartheta $ identifies $\theta= (\theta_1,...,\theta_k)^T$, where  $ \theta_t= E(Y^{(t)}) $ is the mean potential response under treatment $ t $. In the classical exact ANOVA inference, 
the responses are further assumed to have
normal distributions with equal variances. So a common perception is that ANOVA can only be
used for continuous responses. As normality is not necessary in
the asymptotic theory, the ANOVA and the other approaches  introduced next can be used  for non-normal or even discrete responses when $n$ is large. 

To utilize  baseline covariate vector $X$, ANCOVA  is 
 based on the following homogeneous working model, \vspace{-2mm}
\begin{equation}
  E(Y_i \mid A_i,  X_i)  = \vartheta^T A_i  + {\gamma}^T (X_i - \mu_X) ,  \vspace{-2mm} \label{ancova}
\end{equation}
where $\vartheta$
and ${\gamma}$ are unknown vectors having the same dimensions as $A$
and $X$, respectively, and $\mu_X = E(X_i)$.  
 There is no treatment-by-covariate interaction
terms in \eqref{ancova},  which  is  incorrect if patients with different covariates
benefit differently from receiving the same treatment, a scenario that often occurs 
in clinical trials. By Lemma 2 in the supplementary material,  $E\{ Y_i - \vartheta^T A_i - \gamma^T (X_i - \mu_X)\}^2$ is minimized at $(\vartheta, \gamma) = (\theta, \beta) $, where  $\beta =  \sum_{t=1}^k \pi_t \beta_t $ and $\beta_t = \Sigma_X^{-1}{\rm cov}(X_i, Y_i^{(t)})$. Thus, the ANCOVA estimator with working model \eqref{ancova} is model-assisted (Theorems \ref{thm:1} and \ref{theo: CAR2} in \S3). 
Then, what is the impact  of ignoring the 
treatment-by-covariate interaction effect when it actually exists?  
The impact is that the ANCOVA estimator may   be even less efficient than the benchmark ANOVA estimator, as noted by
\citet{Freedman:2008aa} with some examples.

To better adjust for $X$, we consider 
an alternative working model that includes the treatment-by-covariate
interactions:\vspace{-2mm}
\begin{equation}
  E(Y_i\mid A_i , X_i) = \vartheta^T A_i + \sum_{t=1}^k
  \gamma_t^T  (X_i-\mu_X)I(A_i = a_t),\vspace{-2mm}
  \label{hancova}
\end{equation}
where $\vartheta,\gamma_1,\dotsc,\gamma_k$ are unknown  vectors and
$I(\cdot)$ is the indicator function. We call model \eqref{hancova} the  {\em heterogeneous} working model because it includes the interaction terms to accommodate the treatment effect heterogeneity across covariates,    i.e., patients with different 
	covariate values may benefit differently from treatment. By Lemma 2 in the supplementary material, $E\{ Y_i - \vartheta^T A_i - \sum_{t=1}^k	\gamma_t^T  (X_i-\mu_X)I(A_i = a_t) \}^2$  is minimized at $(\vartheta , \gamma_1,...,\gamma_k)=(\theta , \beta_1,...,\beta_k )$, where  $\beta_t = \Sigma_X^{-1}{\rm cov}(X_i, Y_i^{(t)})$, i.e., inference under  working model \eqref{hancova} is also model-assisted.

 To
 differentiate the methods based on \eqref{ancova} and
\eqref{hancova}, we refer to the method based on 
(\ref{ancova}) as ANCOVA and the one based on 
\eqref{hancova} as ANHECOVA.

As a final remark, both  working models \eqref{ancova} and   \eqref{hancova} use the centered covariate
vector $X - \mu_X$. Otherwise, ANCOVA and ANHECOVA do not directly provide estimators of $\theta$. 
Centering is crucial;  
the only non-trivial exception is  when homogeneous
working model \eqref{ancova} is used and linear contrast 
$\theta_t - \theta_s $ is estimated, as the covariate
mean $\mu_X$ cancels out. When fitting the working models \eqref{ancova} and
\eqref{hancova} with real datasets, we can use the least   squares
with $\mu_X$ replaced by $\bar{X}$, the sample mean of   all
$X_i$'s. In other words, we can center the baseline covariates before
fitting the models. Since this step introduces non-negligible
variation to the estimation, it affects the asymptotic variance of
model-assisted estimator of $\theta$ and its estimation for
inference. Thus, we cannot assume the data has been centered in
advance and $\mu_X = 0$ without loss of generality (see \S \ref{sec:
  robust SE}).

\section{Methodology and Theory}
\label{sec: methods}

\subsection{Estimation}
\label{sec:estimation}

We first describe the estimators of $\theta$ under (\ref{anova})-(\ref{hancova}). The ANOVA estimator considered as benchmark is \vspace{-2mm} 
\begin{equation}
  \anova  =
  (\bar{Y}_1,..., \bar{Y}_k )^T, \vspace{-2mm} \label{est1}
\end{equation}
where $\bar{Y}_t$ is the sample mean of the responses $Y_i$'s from
patients under treatment $t$.
As $n \to \infty$, $\anova$ is consistent and asymptotically normal.

Using the homogeneous working model (\ref{ancova}),
the ANCOVA estimator of  $\theta$
is the least squares estimator of the coefficient vector $\vartheta$
in the linear model (\ref{ancova})  with $(A_i,X_i)$ as regressors.  It  has the following
explicit formula,\vspace{-2mm}
\begin{equation}
  \ancova =   \left( \bar{Y}_{1} - \hat\beta^T (\bar{X}_{1} - \bar{X}) , ..., \bar{Y}_k -  \hat\beta^T(\bar{X}_{k} - \bar{X}) \right)^T,\vspace{-2mm}
  \label{est2}
\end{equation}
where $\bar{X}_{t}$ is the sample mean of  $X_i$'s  from patients under treatment $t$, $\bar{X}$ is the sample mean of  all $X_i$'s, and\vspace{-2mm}
\begin{equation}
  \hat{\beta}= \left\{\sum_{t=1}^{k} \, \sum_{ i : A_i =a_t} (X_i - \bar{X}_{t})(X_i - \bar{X}_{t})^{T} \right\}^{-1} \sum_{t=1}^{k} \, \sum_{ i : A_i =a_t} (X_i - \bar{X}_{t})Y_i   \vspace{-2mm}\label{hatbeta}
\end{equation}
is  the least squares estimator of $ \gamma$ in (\ref{ancova}).
It is shown in Theorems 1 and 3 that
$\ancova$ is consistent and asymptotically
normal as $n \to \infty$ regardless of whether  working
model \eqref{ancova} is correct or not, i.e., 
ANCOVA  is model-assisted. 

 The term $ \hat\beta^T(\bar{X}_{t} - \bar{X}) $ in
(\ref{est2}) is an adjustment for covariate $X$ applied to the ANOVA
estimator $\bar{Y}_t$. 
However,  it
may not be the best adjustment in order to reduce the variance. 
A better choice is to use 
heterogeneous working model \eqref{hancova}. The ANHECOVA estimator of
$\theta$ is 
the least squares estimator of $\vartheta$ under model  (\ref{hancova}), \vspace{-2mm}
\begin{equation}
\hancova =   \left( \bar{Y}_{1} - \hat\beta_1^T (\bar{X}_{1} - \bar{X}),  ..., \bar{Y}_k -  \hat\beta_k^T(\bar{X}_{k} - \bar{X})  \right)^T,  \vspace{-2mm}\label{est3}
\end{equation}
where 
\begin{equation}
\hat{\beta}_t =  \left\{ \sum_{ i : A_i =a_t} (X_i - \bar{X}_{t})(X_i - \bar{X}_{t})^{T}
\right\}^{-1}  \sum_{ i : A_i = a_t} (X_i - \bar{X}_{t})Y_i  \vspace{-1mm} \label{hatbetat}
\end{equation}
is the least squares estimator of $\gamma_t $ in (\ref{hancova}) for
each $t$. It is shown in 
Theorems 1-3 below  that the ANHECOVA estimator
$\hancova$ is not only model-assisted, but also asymptotically  at least as efficient as
$\anova$ and $\ancova$, regardless of whether model \eqref{hancova} is
correct or not. 

The following  heuristics reveal why the adjustment  $
\hat\beta_t^T (\bar{X}_t-\bar{X})$ in (\ref{est3}) is better than the adjustment $ \hat\beta^T (\bar{X}_{t} - \bar{X}) $ in (\ref{est2}), and why ANHECOVA often gains but never hurts efficiency even if model (\ref{hancova}) is wrong. 
As the treatment has no effect on $X$, both $\bar{X}_t$ and $\bar{X}$ estimate the same quantity and, hence, $  \hat\beta_t^T (\bar{X}_t-\bar{X}) $ is an ``estimator'' of zero.  As $n \to \infty$,   $\hat\beta_t$ converges to  $\beta_t=
\Sigma_X^{-1} {\rm cov}(X, Y^{(t)}) $ in probability, regardless of
whether (\ref{hancova}) is correct or not (Lemma 3 in the supplementary material). Hence, we can ``replace''  $\hat\beta_t^T (\bar{X}_t-\bar{X}) $  by
$ \beta_t^T (\bar{X}_t-\bar{X})$.  
Under simple  randomization,   \vspace{-2mm}
\begin{equation} \label{eq:greg}
\begin{split}
{\rm var} \{ \bar{Y}_t - \beta_t^T (\bar{X}_t-\bar{X})  \}
& =  {\rm var} (\bar{Y}_t ) + {\rm var} \{ \beta_t^T (\bar{X}_t-\bar{X})\}
-2 {\rm cov} \{ \bar{Y}_t, \beta_t^T (\bar{X}_t-\bar{X})\} \\
& = {\rm var} (\bar{Y}_t ) - {\rm var} \{ \beta_t^T (\bar{X}_t-\bar{X})\}.
\end{split} 
\end{equation}
Consequently, $\bar{Y}_t - \hat\beta_t^T (\bar{X}_t-\bar{X})$ has a
smaller asymptotic variance than $\bar{Y}_t$. 
Note that (\ref{eq:greg}) does not hold with $\beta_t$ replaced by 
other  quantities. This explains why the adjustment 
$\hat{\beta}^T (\bar{X}_t - \bar{X})$ in ANCOVA   may lose  efficiency, as $\hat{\beta}$ in (\ref{hatbeta}) converges to $\pi_1 \beta_1+\cdots +
\pi_k \beta_k$. 

The variance reduction
technique by (\ref{eq:greg}) can be found in the generalized regression (GREG)
approach in the survey sampling literature
\citep{Cassel:1976aa,Sarndal:2003aa,fuller:2009aa,Shao:2014aa,Ta:2020aa}.
From the theory of GREG, $\hat\beta_t$ in \eqref{est3} may be replaced
by any estimator that converges to $\beta_t$  in probability, without affecting the asymptotic distribution of the GREG estimator. This motivates the following potential improvement to \eqref{hatbetat}, which utilizes 
the fact that $X$ has the same covariance across treatments and estimates 
the covariance matrix of $X$ using all patients, 
\begin{equation}
\hat{\beta}_t = \frac{n}{n_{t}} \left\{ \sum_{ i=1}^n  (X_i - \bar{X})(X_i - \bar{X})^{T}
\right\}^{-1}  \sum_{ i : A_i =a_t } (X_i - \bar{X}_t)Y_i  ,  \vspace{-1mm} \label{tildebetat}
\end{equation}
where $n_{t}$ is the number of units under treatment $t$.
This alternative estimator alleviates the concern of using an unstable inverse in (8) when the sample size is small.
In all numerical results in \S4, we apply  \eqref{tildebetat} for ANHECOVA.

\subsection{Asymptotic theory under simple randomization}
\label{sec: inference SR}

We consider asymptotic theory under  simple randomization for a general class of estimators of
the form \vspace{-1mm}
\begin{equation} \label{hattheta}
  \hat \theta (\, \hat b_1,...,\hat b_k) = \left( \bar{Y}_1 -  \hat b_1^T (\bar{X}_1-\bar{X})...,
    \bar{Y}_k - \hat b_k^T (\bar{X}_k-\bar{X}) \right)^T ,
\end{equation}
where $\hat b_t$'s have the same dimension as $X$ and can either be
fixed or depend on the trial data. Note that  class 
\eqref{hattheta} contains all estimators we have
discussed so far:
\begin{equation}
\label{hattheta1}
  \hat \theta (\, \hat b_1,...,\hat b_k) =  \left\{
    \begin{array}{ll}
      \anova & \mbox{ if $\hat b_t =0$ for all $t$} \vspace{1mm} \\
      \ancova & \mbox{ if $\hat b_t =\hat\beta$ in (\ref{hatbeta}) for all $t$} \vspace{1mm} \\
      \hancova & \mbox{ if $\hat b_t =\hat\beta_t $  in (\ref{hatbetat}) or (\ref{tildebetat}) for all $t$} \\
    \end{array} \right.
\end{equation}

\begin{theorem} \label{thm:1}
  Assume (C1) and simple randomization  for treatment assignment.\vspace{-2mm}
  \begin{itemize}
  \item[(i)] Assume that $\hat{b}_t \to b_t$ in probability as $n \to
    \infty$, where $b_t$ is a fixed vector, $t=1,...,k$.
    Then, as $n \to \infty$,
    \begin{equation} \label{normal}
    \sqrt{n} \left\{        \hat \theta (\, \hat b_1,...,\hat b_k) - \theta \right\} \to
    N\left(0, V_{\rm SR}(B)\right) \ \ \mbox{in distribution,} \vspace{-2mm}
    \end{equation}
    where \vspace{-2mm}
    \[
    V_{\rm SR}(B) = {\rm diag}\{\pi_t^{-1} {\rm var}(Y^{(t)} -
    b_t^T X)\} + \mathscr{B}^T \Sigma_X B+ B^T \Sigma_X \mathscr{B}- B^T \Sigma_X B, \vspace{-2mm}
    \]
    ${\rm diag}(d_t)$ denotes the $k \times k$ diagonal matrix with
    the $t$th diagonal element $d_t$, 
    $\mathscr{B} = (\beta_1,...,\beta_k)$, the matrix with columns 
    $\beta_1,...,\beta_k$, and $B = (b_1,...,b_k)$.
    In particular, (\ref{normal}) holds for $\anova$, $\ancova$, and $\hancova$  as described by (\ref{hattheta1}). \vspace{-2mm}
 \item[(ii)] (Optimality of ANHECOVA). $V_{\rm SR}(B)$ is minimized at $B =\mathscr{B}$ in the sense that
    $V_{\rm SR}(B) - V_{\rm SR}(\mathscr
    {B})$ is  positive semidefinite for all $B$.
  \end{itemize}
\end{theorem}
 We briefly describe the proof for part (ii) in Theorem 1 and defer other details to the supplementary material. Notice that
  \begin{equation}
	V_{\rm SR}(B)- V_{\rm SR}(\mathscr{B})=  \text{diag}\{\pi_t^{-1} (\beta_t - b_t)^T \Sigma_X(\beta_t - b_t) \} -  (\mathscr{B}- B)^T\Sigma_X(\mathscr{B}- B). \nonumber
\end{equation}
 The positive semidefiniteness of this matrix follows from the following algebraic result with $ M= \Sigma_X^{1/2} (\mathscr{B}- B) $.

\begin{lemma} \label{lem:1}
 Let $M$ be a matrix whose columns are $m_1,...,m_k$, and $\pi_1,...,\pi_k$ be nonnegative constants with $\sum_{t=1}^k \pi_t =1$. Then 
${\rm diag} ( \pi_t^{-1} m_t^T m_t) - M^T M$ is  positive semidefinite. 
\end{lemma}

We would like to emphasize that \Cref{thm:1}(i) holds regardless of
whether model (\ref{hancova}) is correct or not. 
\Cref{thm:1}(ii) shows that ANHECOVA not only has guaranteed efficiency gain over ANOVA, but is
also the most efficient estimator within the class of estimators
in \eqref{hattheta}  as it  attains the  optimal $ V_{\rm SR}(\mathscr{B}) $. 
Another consequence of \Cref{thm:1}(ii) is that adjusting for more
covariates in ANHECOVA does not lose and often gains asymptotic
efficiency, although adjusting for fewer covariates may
have better performance when $n$ is small.

For the important scenario of estimating the linear contrast $\theta_t
- \theta_s$ with fixed $t$ and $s$, the corresponding model-assisted
estimator is $  c_{ts}^T \hat \theta$, where $\hat \theta$ is given by
\eqref{hattheta} and $c_{ts}$ is the $k$-dimensional vector whose
$t$th component is 1, $s$th component is $-1$, and other components
are 0. The following corollary provides an explicit comparison of the
asymptotic variances of ANOVA, ANCOVA, and ANHECOVA estimators of linear contrasts, showing that 
the ANHECOVA estimator has strictly smallest asymptotic variance except for  some very special cases.

\begin{corollary} \label{coro1}
  Assume (C1) and simple randomization.
  \begin{itemize}
  \item[(i)] For any $t$ and $s$, 
    the difference between the asymptotic variances of $  \sqrt{n} c_{ts}^T ( \anova - \theta )$ and $ \sqrt{n}  c_{ts}^T ( \shancova -\theta )$ is
    $$  \frac{(\pi_s \beta_t + \pi_t \beta_s)^T \Sigma_X  (\pi_s \beta_t + \pi_t \beta_s)}{\pi_t \pi_s (\pi_t + \pi_s)}
    + \frac{ (1-\pi_t - \pi_s) (\beta_t - \beta_s)^T \Sigma_X (\beta_t-\beta_s)}{\pi_t + \pi_s}, $$
    which is always $\geq 0$ with equality holds  if and only if
    \begin{equation}
      \pi_s \beta_t + \pi_t \beta_s = 0 \quad \mbox{and} \quad
      (\beta_t -\beta_s ) (1- \pi_t - \pi_s ) = 0.  \label{c1}
    \end{equation}
  \item[(ii)]  For any $t$ and $s$, the difference between the asymptotic variances of $\sqrt{n} 
    c_{ts}^T( \ancova - \theta )$ and $  \sqrt{n}  c_{ts}^T (\shancova - \theta ) $ is
    $$  \frac{ (\beta_t -  \beta)^T \Sigma_X (\beta_t-
      \beta)}{\pi_t}+ \frac{ (\beta_s - \beta)^T  \Sigma_X
      (\beta_s-\beta)}{\pi_s}   -  (\beta_t -  \beta_s)^T \Sigma_X
    (\beta_t- \beta_s),
    $$
    which is always $\geq 0$ with equality holds  if and only if
    \begin{equation}
       \beta= \frac{\pi_s \beta_t + \pi_t \beta_s}{\pi_t + \pi_s}
      \quad \mbox{and} \quad
      (\beta_t -\beta_s ) (1- \pi_t - \pi_s ) = 0 . \label{c2}
    \end{equation}
  \end{itemize}
\end{corollary}

When $k=2$, i.e.,  there are only two arms,  (\ref{c1}) reduces to
$\pi_2\beta_1 + \pi_1 \beta_2 = 0$, and (\ref{c2}) reduces to $\beta_1 =
\beta_2$ or $\pi_1 = \pi_2 = 1/2$. The same conclusions were also
obtained by \cite{Lin:2013aa} under a different framework that only
considers the randomness in  treatment assignments.
\cite{Liu:2020aa} extended the results in  \cite{Lin:2013aa} to
stratified simple randomization. 

We would like to emphasize that when there are more than two arms ($k >
2$), \eqref{c1} or \eqref{c2} only holds in denegerate or peculiar
cases. For the comparison of ANHECOVA with ANOVA, (\ref{c1}) holds if and only
if $\beta_t = \beta_s =0$, because $0< \pi_t + \pi_s < 1$ when
$k>2$. For the comparison of ANHECOVA with ANCOVA, (\ref{c2}) holds if
and only if $\beta_t = \beta_s = {\beta} = \sum_{t=1}^k
\pi_t\beta_t$. Therefore, $\beta_t = \beta_s$ is not enough for
ANCOVA to be as efficient as ANHECOVA for estimating $\theta_t - \theta_s$.
Moreover, even if treatment allocation is balanced, i.e., $\pi_1 =
\cdots = \pi_k$, ANCOVA is generally less efficient than ANHECOVA when
there are more than two arms; this is different from the conclusion in
the case of two arms. Finally, in estimating $\theta_t -
\theta_s$ for \emph{all pairs} of $t$ and $s$, for ANOVA to have
the same asymptotic efficiency as ANHECOVA, all $\beta_t$'s need to
be zero, i.e., $X$ is uncorrelated with $Y^{(t)}$ for every $t$;  
for ANCOVA to have the same asymptotic efficiency as  ANHECOVA, all $\beta_t$'s must be the same, i.e., models (\ref{ancova}) and  (\ref{hancova}) are the same. 

It is worth to mention that when there are more than two treatment arms, the  ANCOVA estimator can be either more efficient or less efficient than the ANOVA estimator even under balanced treatment allocation. This is also observed by \cite{Freedman:2008aa} in some specific examples.

	The asymptotic equivalence between ANCOVA and ANHECOVA in the special 
	scenario of considering a linear contrast under two arms with equal allocation 
 has led to 
	an  imprecise  recommendation of ANCOVA over ANHECOVA  under this circumstance \citep{Wang:2019aa, ma2020regression}. 
In addition to the previous discussion about the inferiority of ANCOVA  for linear contrasts under either multiple arms or unbalanced treatment allocation, it follows from  
 Theorem \ref{thm:1} that inference based on ANHECOVA is in general asymptotically more efficient 
	than that based on ANCOVA when  functions of $ \theta $ other than linear contrasts
are concerned (such as a ratio or an odds ratio based on two components of $\theta$)
	even in the case of  two arms with equal treatment allocation.

\subsection{Asymptotic theory under covariate-adaptive randomization}
\label{sec: inference CAR}

We now consider the estimation of $\theta$ under  covariate-adaptive randomization  as described in \S2.2. Specifically, we would like to provide answers to the following two questions: 
Is there an estimator achieving  wide applicability  and universality, i.e., 
the estimator has an asymptotic distribution  invariant with respect to all commonly used randomization schemes so that the same inference
procedure can be constructed regardless of which randomization scheme is
used? 
Is there an estimator that is asymptotically the most efficient within the class of estimators given by  (\ref{hattheta}) under any covariate-adaptive  randomization?

The answers  to these two questions are affirmative, as established formally
in Theorems \ref{theo: CAR1} and \ref{theo: CAR2}, respectively.  Importantly, the key to achieve wide applicability  and universality as well as guaranteed efficiency gain is using the ANHECOVA estimator  $\hancova$ 
with {\em all the joint levels of $Z $ included in the covariate $
X $}. 

\begin{theorem}\label{theo: CAR1}
	(Wide applicability and Universality of ANHECOVA).  Assume (C1) and (C2).
	If heterogeneous model (\ref{hancova}) is used  with $X$  containing the dummy variables for all the joint levels of $ Z $ as a  sub-vector,
	then,   regardless of whether working model (\ref{hancova}) is correct  or not and which randomization scheme is used, as $ n\rightarrow\infty $,
  \begin{equation} \label{normal1}
    \sqrt{n} \left(\shancova - \theta \right) \to
    N\big(0, V_{\rm SR}(\mathscr{B})\big) \ \ \mbox{in distribution},
  \end{equation}
  where $V_{\rm SR}(\mathscr{B}) ={\rm diag}\{\pi_t^{-1} {\rm var}(Y^{(t)}
  - \beta_t^T X)\} + \mathscr{B}^T \Sigma_X \mathscr{B}$ and $\mathscr{B}= ( \beta_1,...,\beta_k)$. 
\end{theorem}

Comparing \Cref{thm:1} with \Cref{theo: CAR1}, we see that the ANHECOVA
estimator including all dummy variables for $Z$ has exactly the same
asymptotic variance in simple randomization and any covariate-adaptive
randomization satisfying (C2), which is reflected by the fact that 
$V_{\rm SR}(\mathscr{B})$ is the same as $V_{\rm SR}(B)$ in (\ref{normal}) with $B= \mathscr{B}$. Therefore, this estimator achieves wide  applicability and universality. 
As we show next, however, this is not true  for ANOVA or 
ANCOVA using  model \eqref{ancova}, or for ANHECOVA when $Z$ is not fully included in the working model. 

 To answer the second question,  we need a further condition on the randomization scheme, mainly for estimators not using model (\ref{hancova}) or not including all levels of $Z$ in $X$.
\begin{itemize}
	\item[(C3)] 
There exist
	$ k\times k $ matrices $\Omega(z) $, $ z\in \mathcal{Z} $, such that, as
	$n\rightarrow\infty$,
	\[
	\sqrt{n} \left(      {\textstyle \frac{n_1(z) }{n(z)}- \pi_1 , \ldots , \frac{n_k (z) }{n(z)}- \pi_k  }, \, z\in \mathcal{Z} \right)^T \mid Z_1, \dots, Z_n\to  N \left(0, D \right) 	\ \ \mbox{in distribution,}  
   \]
 where $ D $ is a block diagonal matrix whose blocks are  matrices $ \Omega(z)/P(Z_i=z), z\in \mathcal{Z} $.
\end{itemize}

Condition (C3) weakens Assumption
4.1(c) of \cite{Bugni:2019aa} in which $\Omega(z)$ takes a more special
form. For simple randomization, $\Omega(z) = {\rm diag} ( \pi_t)- \pi \pi^T $  for all $
	z $, where $ \pi=
	(\pi_1,\dots, \pi_k)^T$.  For stratified permuted block randomization and
	stratified biased coin randomization, $\Omega(z) = 0$ for all $z$. Note that 
	Pocock-Simon's minimization  scheme does not satisfy (C3) because the treatment assignments are correlated across strata,
	although some recent theoretical result has been obtained  
	\citep{hu2020theory}.
	Thus, the following result does not apply to Pocock-Simon's minimization. 
	However, our \Cref{theo: CAR1} applies to
	minimization, as  (C3) is not needed in Theorem 2.

The next theorem establishes the asymptotic distributions of estimators in  class (\ref{hattheta}) under covariate-adaptive randomization, based on which we show the optimality of the ANHECOVA estimator.
\begin{theorem}\label{theo: CAR2}
	Assume (C1), (C2), and (C3). Consider class (\ref{hattheta}) of estimators and, without loss of generality, we assume that all levels of $Z$ are included in $X$, as the components of $\hat b_t$'s in (\ref{hattheta}) corresponding to levels of $Z$ not in $X$ may be  set to 0. \vspace{-2mm}
  \begin{itemize}
  	\item[(i)]     For $\hat \theta (\, \hat b_1,...,\hat b_k) $ defined in
  	(\ref{hattheta}) with $\hat{b}_t \to b_t$ in probability as $n \to
  	\infty$,  $t=1,...,k$, \vspace{-2mm}
  	\begin{equation} \label{normal2}
  	\sqrt{n} \left\{        \hat \theta (\, \hat b_1,...,\hat b_k) - \theta \right\} \to
  	N\big(0, V(B)  \big)\quad \text{in distribution},\vspace{-2mm}
  	\end{equation}
  	where \vspace{-2mm}
\begin{equation}
  	V(B)= V_{\rm SR}(B) - E\left[ R(B) \{  \Omega_{\rm SR}- \Omega (Z_i) \} R(B)\right], \label{rb}\vspace{-2mm}
\end{equation}
  	$V_{\rm SR}(B)$ is given in (\ref{normal}), 
 $B = (b_1,...,b_k)$, 
  	$\Omega_{\rm SR} = {\rm diag} ( \pi_t)- \pi \pi^T $, and
  	$R(B) = {\rm diag} \big( \pi_t^{-1} E\{Y_i^{(t)}-\theta_t -  b_t^T (X_i-\mu_X)\mid
  	Z_i\}\big) $. Furthermore, $R(\mathscr{B} )= 0$ and, hence, 
  	$V(\mathscr{B})= V_{\rm SR}(\mathscr{B})$, where $\mathscr{B}
  	=   (\beta_1,...,\beta_k)$.  \vspace{-2mm}
  \item[(ii)]  (Optimality of ANHECOVA).  $V(B)$ is minimized at $B =\mathscr{B}
 $ in the sense that
  $V(B) - V(\mathscr
  {B})$ is positive semidefinite for all $B$. 
  \end{itemize}
\end{theorem}

The main technical challenge in the proofs of \Cref{theo: CAR1} and
\Cref{theo: CAR2} is that the treatment assignments
$A_1,\dotsc,A_n$ are not independent due to covariate-adaptive randomization, so we cannot directly apply the
classical Linderberg central limit theorem. Instead, we  decompose $   \hat \theta (\, \hat b_1,...,\hat b_k) - \theta  $ into four terms and then apply a conditional version of the Linderberg central limit theorem to handle the dependence.  The details  can be found in the supplementary material.  
	
A number of conclusions can be made from Theorem \ref{theo: CAR2}. 
\vspace{-2mm}
\begin{enumerate}
	\item 
With Theorem 2 answering the first question in the beginning of \S3.3, i.e., $\hancova$ with all joint levels of $Z$ included in model (\ref{hancova}) achieves wide applicability and universality, the second question is  answered by Theorem 3(ii) showing that $\hancova$ is asymptotically the most efficient estimator compared with all estimators in  class (\ref{hattheta}); in particular, $ \hancova $ attains guaranteed efficiency gain under any covariate-adaptive randomization satisfying (C2). 	 Our optimality result in Theorem \ref{theo: CAR2}(ii) is about the joint estimation of the vector $\theta$, which is 
	substantially more general than the existing one-dimensional optimality results  about linear contrasts. Furthermore,  our conclusion made in \S3.2, i.e.,    ANHECOVA is asymptotically superior  over ANCOVA except for the particular case of estimating a linear contrast for two arms with balanced treatment allocation, holds for all commonly used covariate-adaptive  \vspace{-2mm} randomization schemes. 
	\item 
	A price paid for not using model (\ref{hancova}) or not including all levels of $Z$ in (\ref{hancova}) is that the asymptotic validity of  the resulting estimator
	requires condition (C3),    which  is not needed in Theorem \ref{theo: CAR1}.
 Furthermore, the resulting estimator not only is less efficient according to the previous conclusion, but also  has a more complicated asymptotic covariance matrix depending on the randomization schemes (universality is not
 satisfied), which requires extra handling in variance estimation for inference; see, for example, \cite{Shao:2010aa}, 
 \cite{Bugni:2017aa}, and \cite{Ma:2020aa}.
 \vspace{-2mm}
	\item
	 Under covariate-adaptive randomization satisfying (C2)-(C3), it is still true that the ANCOVA estimator  using  model \eqref{ancova} may be asymptotically more efficient or less efficient than the benchmark ANOVA estimator. \vspace{-2mm}
	 \item From (\ref{rb}), the asymptotic covariance matrix $V(B)$ is invariant with respect to randomization scheme if   $R(B)$ in (\ref{rb}) is 0, which is the case when $B= \mathscr{B}$, i.e., $\hancova$ is used with all levels of $Z$ included in $X$.
If $R(B)$ is not 0, such as the case of ANOVA, ANCOVA, or ANHECOVA not adjusting for all joint levels of $ Z $, then  $V(B)$ depends on randomization scheme and, the smaller the $\Omega(z)$, the
more efficient the estimator is. Thus, the stratified permuted block or biased coin with $\Omega (z) =0$ for all $z$ is preferred in this regard. 
\vspace{-2mm}
	 \item  The roles played by design and modeling can be understood through
	 \vspace{-2mm}
	 $$ 	V(B)- V_{\rm SR}(0)= \{V_{\rm SR}(B)- V_{\rm SR}(0) \}- E\left[ R(B) \{  \Omega_{\rm SR}- \Omega (Z_i) \} R(B)\}\right], \vspace{-2mm}$$ 
	 where $ V_{\rm SR}(0) $ is the asymptotic variance of ANOVA estimator under simple randomization. As we vary the randomization scheme and the working model, the change in the asymptotic variance is determined by two terms. The first term $   \{V_{\rm SR}(B)- V_{\rm SR}(0) \}$ arises from using a working model; the second term $ E\left[ R(B) \{  \Omega_{\rm SR}- \Omega (Z_i) \} R(B)\right] $ is the reduction due to using a covariate-adaptive randomization scheme, which also depends on  the working model being used via $ R(B) $. Therefore, it is interesting to note that although the primary reason of using covariate-adaptive randomization is to achieve balance of treatment groups across prognostic factors, it also improves statistical efficiency. \vspace{-2mm}
\end{enumerate}

 Theorem 3 together with a further derivation leads to the following result. 

\begin{corollary}[Duality between design and analysis] \label{coro2}
	Assume (C1)-(C3) and that  $ X $ only includes the dummy variables for all joint levels of $Z$. Then, for any $B$ in (\ref{normal2}), 
	$V(B) = V_{\rm SR}( \mathscr{B})  + E \{ R(B) \Omega(Z_i) R(B)\}.$
\end{corollary}

A direct consequence from Corollary 2 is that, if $\Omega(z) = 0$ for all $z$ (e.g., stratified permuted block or biased coin randomization is used) and  
$X$ \emph{only} includes all joint levels of $ Z $, then all estimators in class \eqref{hattheta}, including the benchmark ANOVA estimator,  have the same asymptotic efficiency as the ANHECOVA estimator under any randomization. This shows the duality between design and analysis, i.e., modeling with all joint levels of $ Z $ is equivalent to designing with $Z$.

\subsection{Robust standard error}
\label{sec: robust SE}

For model-assisted inference on a function of $\theta$ based on 
Theorems 1-3, a crucial step is to construct a consistent
estimator of asymptotic variance. 
The
customary linear model-based variance estimation assuming
homoscedasticity can be inconsistent, as criticized by
\cite{Freedman:2008aa}  and \cite{fda:2019aa}. Therefore, it is important that we use variance estimators that are consistent regardless of whether the working model is correct or not and whether heteroscedasticity is present or not. 

Consider the ANHECOVA estimator $\hancova$ in \eqref{est3} using either (\ref{hatbetat}) or \eqref{tildebetat}, 
where covariate $X$ includes all dummy variables for
$Z$ that is used in the randomization. There exist formulas for
heteroscedasticity-robust standard error (such as those
provided in the \texttt{sandwich} package in
\texttt{R}). However, those formulas cannot be directly
applied here, because they do not account for the additional variation
introduced by centering the covariate $ X $ as required by the
identification of $ \theta $.  In fact,
the term  $\mathscr{B}^T \Sigma_X \mathscr{B}$ in the asymptotic
variance $V_{\rm SR}(\mathscr{B})$ in \Cref{theo: CAR1} arises from centering
$X$.

Instead, we should use the robust variance estimator based on
$V_{\rm SR}(\mathscr{B})$, as described next. Let $\hat \Sigma_X$ be the sample
covariance matrix of $X_i$ based on the entire sample and
$S_t^2(\hat\beta_t)$ be the sample variance of $(Y_i - \hat\beta_t^T
X_i)$'s based on the patients in treatment arm $t$. Then
$V_{\rm SR}(\mathscr{B})$ in  (\ref{normal1}) can be estimated by \vspace{-2mm}
\begin{equation}
  \label{varest}
  \hat{V} = \text{diag}\{\pi_t^{-1} S_t^2(\hat\beta_t)\} +
  \hat{\mathscr{B}}^T \hat\Sigma_X \hat{\mathscr{B}},\vspace{-2mm}
\end{equation}
where $\hat {\mathscr{B}}$ is ${\mathscr{B}}$ with $\beta_t$ replaced by $\hat\beta_t$. 
This variance estimator is consistent as $n \to \infty$ regardless of
whether the heterogeneous working model \eqref{hancova} or 
homoscedasticity  holds or not, and regardless of which
randomization scheme is used. 

In many applications the primary analysis is about
treatment effects in terms of the linear contrast $\theta_t -
\theta_s = c_{ts}^T \theta  $ for one or several pairs of $(t,s)$. For
large $n$, an asymptotic level $(1-\alpha)$ confidence interval of
$\theta_t - \theta_s$ is\vspace{-2mm}
$$\left(c_{ts}^T \shancova - z_{\alpha /2}{\rm SE}_{ts},  \
  c_{ts}^T \shancova +  z_{\alpha /2} {\rm SE}_{ts}\right), \vspace{-2mm}$$
where ${\rm SE}_{ts}^2 = \pi_t^{-1}     S_t^2(\hat\beta_t) +\pi_s^{-1}  S_s^2(\hat\beta_s) +  (\hat\beta_t-\hat\beta_s)^T \hat\Sigma_X (\hat\beta_t   - \hat\beta_s)$ and
$z_\alpha$ is the $(1-\alpha)$ quantile of the standard normal distribution.
The same form of confidence interval can be used for any linear contrast $c^T\theta$ (the sum of components of $c$ is 0)
with $c_{ts}^T \shancova$ and ${\rm SE}_{ts}^2$ replaced by $c^T \shancova$
and ${\rm SE}_c^2 = c^T \hat V c$, respectively.
Let ${\mC}$ be the collection of all linear contrasts with dimension $k$.
An asymptotic level $(1-\alpha)$ simultaneous confidence band of
$c^T\theta$, $c \in {\mC}$, can be obtained by Scheff\'{e}'s method, \vspace{-2mm}
$$\left(c^T \shancova - \chi_{\alpha,
    k-1} \, {\rm SE}_c,  \
  c^T \shancova +  \chi_{\alpha, k-1} \, {\rm SE}_c \right), \ \ c \in \mC ,\vspace{-2mm}$$
where  $\chi_{\alpha, k-1}$ is the square root of the $(1-\alpha)$ quantile of the
chi-square distribution with $(k-1)$ degrees of
freedom.  Correspondingly, to test the hypothesis 
$H_0 : \theta_1=\cdots =
\theta_k$, an asymptotic level $\alpha$ chi-square test rejects $H_0$
if and only if \vspace{-2mm}
$$  \shancova^T C^T(C \hat{V} C^T)^{-1} C \shancova >
\chi^2_{\alpha,k-1}, \vspace{-2mm}$$ 
where $C$ is the $(k-1)\times k$ matrix whose
$t$th row is $c_{tk}^T$, $t=1,...,k-1$.

Inference procedures based on  the ANOVA or ANCOVA estimator can be
similarly obtained using Theorems 1 and 3. However,
as they do not achieve universality,  a tailored derivation
is needed for each covariate-adaptive randomization scheme. 
For example,  under  the stratified permuted block or biased coin randomization, the ANOVA or ANCOVA estimator is asymptotically more efficient than the same estimator under simple randomization; thus, 
 using variance estimators valid only under simple randomization may lead  to unduly conservative  inference \citep{fda:2019aa}. To eliminate the conservativeness, modifications depending on covariate-adaptive randomization schemes have to be made \citep{Shao:2010aa,Bugni:2017aa}. For Pocock-Simon's minimization, however, how to derive the tailored variance estimators for the ANOVA and ANCOVA estimators is  not yet known as the asymptotic properties of the minimization scheme is still not well established.
This is why we recommend ANHECOVA over the other model-assisted
estimators for the practice.

\section{Empirical Results}

\subsection{Simulation results}

We perform a simulation study based on the placebo arm of 481 patients
in a real clinical trial to demonstrate the 
finite-sample properties of the model-assisted procedures. We use
the observed continuous response of these 481 patients as the
potential response $Y^{(1)}$ under treatment arm 1, and a 
2-dimensional continuous baseline covariate $(U,W)$. 
The empirical distribution of $(Y^{(1)},U,W)$ of these  patients is the population distribution in simulations.
Notice that we do not know the true relationship
between $Y^{(1)}$ and $(U,W)$ because they are from the real data. 
 In fact, a linear model fit between $Y^{(1)}$ and $(U,W)$ based on 481 patients results in multiple and adjusted R-squares $\leq 0.05$.  Thus,
working models \eqref{ancova} and \eqref{hancova} are likely to be 
misspecified in our simulation.

We consider two simulation settings that are different in how the potential responses $Y^{(2)}$ and
$Y^{(3)}$ of the other two treatment arms are generated, and how the
treatment assignment is randomized.  
Our first simulation compares the standard deviations of the
ANOVA, ANCOVA, and ANHECOVA estimators  of $\theta_2
-\theta_1$, with $X = U$ for ANCOVA and ANHECOVA. 
The two additional
potential responses are generated according to
\begin{align}
	\begin{split}
			Y^{(2)} & =  Y^{(1)}+\zeta (U-\mu_U)\\
	Y^{(3)} & = Y^{(2)} \label{eq: simu1}
	\end{split}
\end{align}
($\theta_1 = \theta_2 = \theta_3$, i.e., no average treatment effect). 
The sample size is 481 (all data points are sampled). Treatments are
assigned by simple randomization  according to three different allocation
proportions: 1:2:2, 1:1:1, and 2:1:1. Thus, the only randomness in the first simulation setting is from treatment assignments.  
Here $\mu_U$ is the mean of 481
$U$-values. Since $\beta_t = \Sigma_X^{-1}{\rm cov}(X_i, Y_i^{(t)})$, 
$ \beta_2=\beta_3 = \beta_1 + \zeta$. The value of  $\beta_1 $ is $ - 0.255$ and 
the value of  $\zeta$ represents the  magnitude of treatment-by-covariate interaction. But $\zeta$ does not affect the average treatment effect as it is the coefficient in front of  centered $U- \mu_U$.
Although we only consider
the estimation of $\theta_2 -\theta_1$, data from the third arm is
still used by ANCOVA and  ANHECOVA.

Based on 10,000 simulations, all three estimators have negligible
biases and their standard deviations are plotted  in Figure 1
for different values
of $\zeta$ between $-1$ and 1. The simulation result shows that, as predicted by
our theory, ANHECOVA is generally more efficient than the other two estimators,
except when $\zeta$ is nearly $0$ where ANCOVA is comparable to
ANHECOVA. Furthermore, the simulation with allocation 1:2:2  (left
panel in Figure 1) shows very clearly that there is no definite ordering of
the variances of ANCOVA and ANOVA. Our \Cref{coro1} suggests that a
balanced allocation does not guarantee the
superiority of ANCOVA over ANOVA when there
are multiple
arms (in contrast with the case of two arms), which can be seen from the simulations with allocation 1:1:1 (middle panel in Figure 1).

The second simulation setting is intended to examine the
performance of  estimators,  standard errors, and the proposed 
 95\% asymptotic confidence intervals for $\theta_2
-\theta_1$ and $\theta_3-\theta_1$, under three randomizations schemes, simple randomization, stratified permuted block, and Pocock-Simon's minimization,
with allocation 1:1:1 or 1:2:2. For each simulation, a random sample of  size $n=200$
or 400 is drawn from the 481 subjects' $(Y^{(1)},U,W)$ with
replacement, and $Y^{(2)}$ and $Y^{(3)}$ are generated according to 
\begin{align}
	\begin{split}
	Y^{(2)} & =  -1.3 + Y^{(1)}- 0.5 (U-\mu_U) -0.01 (U^2 - \mu_{U^2}) + 0.3 (W-\mu_W) \\
Y^{(3)} & = -1 + Y^{(1)} - 0.1(U-\mu_U) -0.01 (U^2 - \mu_{U^2}) -0.1  (W-\mu_W)  \label{eq: simu2}
	\end{split}
\end{align}
(average treatment effects $\theta_2 -  \theta_1 = -1.3$ and $\theta_3 -   \theta_1 =-1$).
The magnitude of treatment-by-covariate interaction is represented by the differences of $\beta_t$-values, where for $ X= (U, W)^T $, 
$\beta_1 = (-0.240, -0.001)^T$, 
$\beta_2 = (-0.853, 0.298)^T$, and $\beta_3 = (-0.453, -0.102)^T$.  
Note that a quadratic term  $U^2 - \mu_{U^2}$ appears in the data generating process \eqref{eq: simu2} but is not adjusted by ANCOVA or ANHECOVA. Thus, the models for $Y^{(2)}-Y^{(1)}$ and 
$Y^{(3)}-Y^{(1)}$ are misspecified, in addition to the likely event that the model for $Y^{(1)}$ is misspecified.

The covariate $Z$ used in randomization is composed of  a three-level
discretized $W$ (with proportions 0.24, 0.22, and 0.54) and a two-level discretized $U$ (with proportions 0.77 and 0.23). These $Z$-categories are created according to the disease severity encoded by covariates $U$ and $W$. For stratified permuted block randomization,   block size  6 is used under 
treatment allocation 1:1:1 and  block size 10 is used under treatment allocation
1:2:2. 

For ANCOVA and
ANHECOVA, we consider two working models with different choices of
$X$. One model includes the dummy variables for $Z$  but not $(U,W)$,
motivated by the fact that $Z$ is a discretization of $(U,W)$. The other model includes not only the dummy variables for
$Z$, but also $U$ and $W$. The simulation results with $ n=200 $ and $ n=400 $ based on
10,000 simulations are shown in Tables \ref{tb:200} and \ref{tb:400} respectively. 

Note that in the second simulation when covariate-adaptive
randomization is used, for ANOVA or ANCOVA, we employ the standard
error derived under
simple randomization based on \Cref{thm:1}.  According to our theory, it is
expected that the standard errors and the related confidence intervals based on ANOVA and ANCOVA are conservative; the simulation can show how serious the conservativeness is. 

The following is a summary of simulation results in Tables \ref{tb:200} and \ref{tb:400}. \vspace{-2mm}
\begin{enumerate}
	\item
	All estimators have negligible bias compared to their standard
	deviation.\vspace{-2mm}
	\item ANHECOVA has the smallest standard deviation in all scenarios of our
	simulation. This is consistent with our asymptotic theory.\vspace{-2mm}
	\item There is no unambiguous ordering of
		the standard deviations of ANCOVA and ANOVA.  In particular,
	ANCOVA  is better in estimating $\theta_2 -\theta_1$ but worse in estimating 
	$\theta_3-\theta_1$.  \vspace{-2mm} 
	\item For ANHECOVA,  including
	additional covariates $U$ and $ W$ in the working model results in a smaller
	standard deviation,  indicating that $U$ and $W$ carry more useful information than their discretized values. Interestingly, this is not always the case for ANCOVA.\vspace{-2mm}
	\item  From Tables 1 and 2, the performances of ANHECOVA are nearly the same under three randomization schemes,  the simple randomization, stratified permuted block, and Pocock-Simon's minimization. This supports the universality results in our asymptotic  theory. \vspace{-2mm}
		\item Under simple randomization, the robust standard errors for all  model-assisted estimators are very
	close to their actual standard deviations, and  confidence intervals have
	 nominal coverage in all settings. However, although this is still true for ANHECOVA under stratified permuted block and Pocock-Simon's minimization, it  is not the case for ANOVA and ANCOVA, i.e., 
	standard errors valid under simple randomization
	appear to overestimate the actual standard deviations, so the
	confidence intervals are conservative. This observation reflects the universality property of ANHECOVA. 
\end{enumerate}

\subsection{A real data example}

We further illustrate the different model-assisted procedures using a
real data example. \cite{Chong:2016aa} conducted a randomized
experiment to evaluate the impact of low dietary iron intake on human
capital attainment. They recruited
students of age 11 to 19 in a rural area of Cajamarca, Peru, where
many adolescents suffer from iron deficiency. The goal of this
randomized trial is to quantify the
causal effect of reduced adolescent anemia on school attainment. By
using students' school grade as covariate $ Z $ with five levels, a
stratified permuted block randomization with 1:1:1 allocation was
applied to assign 219 students to one of the following three
promotional videos: \vspace{-2mm}
\begin{description}
	\item[Video 1:] A popular soccer player is encouraging iron supplements to maximize energy;\vspace{-3mm}
	\item[Video 2:] A physician is encouraging iron supplements for overall health;\vspace{-3mm}
	\item[Video 3:] A dentist encouraging oral hygiene without 
	mentioning iron at all.\vspace{-1mm}
\end{description}
\cite{Chong:2016aa} investigated whether showing different promotional
videos to the students can improve their academic performance through
increased iron intake. Video 3 is treated as a ``placebo''.
After the treatment assignments, four students were excluded from the
analysis for various reasons, which we also ignore in our
analysis. The dataset is available at
\texttt{https://www.openicpsr.org/openicpsr/project/113624/version/V1/view}.

\cite{Chong:2016aa} used various outcomes in their analysis; here we
focus on one of their primary outcomes---the academic
achievement---as an example. In this trial, the academic achievement
is measured by
a standardized average of the student's grades in math, foreign
language, social sciences, science, and communications in a
semester. For the model-assisted approaches, we use the baseline
anemia status as the covariate in  working models \eqref{ancova}
and \eqref{hancova}, which is believed to moderate the treatment
effect \citep{Chong:2016aa}.

Table 3 reports the analysis results by using different model-assisted
procedures. Like in our simulation studies, the standard
errors for ANOVA and ANCOVA are computed using estimator based on
\Cref{thm:1} for simple randomization, even though the randomization
scheme here is covariate-adaptive. All the model-assisted estimators
find similar effect sizes for the two contrasts (physician versus
placebo, soccer star versus placebo), and the two ANHECOVA estimators have
slightly smaller standard errors. Including baseline anemia status in
the working model are useful to reduce the standard
error. Compared to the placebo, the promotional video
by the soccer player does not seem to have a positive effect on the
academic achievement. In contrast, the video of the physician
promoting iron supplements appears to have a positive effect.
The difference between ANHECOVA and ANOVA or ANCOVA, and between including and not including anemia 
	can be seen from the magnitude of the corresponding p-values. 
\vspace{-5mm}

\section{Recommendation and Discussion}
\label{sec: discussion}

To improve its credibility and efficiency, we believe a clinical trial analysis can benefit from  considerations outlined in
\S1.1 and discussed throughout \S2-3.   

Our theoretical investigation shows that 
the ANHECOVA  with  all
joint levels of $Z$ included in heterogeneous working model (\ref{hancova}),  coupled  with the robust variance estimator given by  (\ref{varest}), achieves guaranteed efficiency gain over benchmark ANOVA, asymptotic optimality among a large class of estimators, wide applicability and universality. 
Thus, we believe it deserves wider usage in the clinical
trial practice. In addition to all joint levels of $ Z $, other  baseline covariates highly associated with the responses can also be 	
included  in the ANHECOVA working model, following the guidance of 
\cite{fda:2019aa}. 
Our theory shows that using 
ANOVA, ANCOVA  with  model (\ref{ancova}), or ANHECOVA not adjusting for all joint levels of $ Z $, suffers  from invalidity, inefficiency, or non-universality in the sense that the asymptotic distribution of the estimator depends on a particular randomization scheme.

Our model-assisted  asymptotic theory is given in terms of the joint asymptotic distribution in estimating $\theta$, the vector of mean responses, with multiple treatment arms 
under a wide range of covariate-adaptive randomization schemes. 
It can be readily used for inference about linear or nonlinear functions of $\theta$, with either continuous or
discrete responses. Although  working models \eqref{ancova}
and \eqref{hancova} are not commonly used for discrete responses, ANHECOVA is still asymptotically valid as it is model-assisted. 
For binary responses, a popular  model is
logistic regression. However, if $X$ has a continuous
component, the standard logistic regression inference is model-based
instead of model-assisted and, thus, it may be invalid 
if the logistic model is not correctly specified.  It should also be noted that 
the standard logistic regression attempts to estimate a conditional treatment effect,
which is distinct from the unconditional treatment effect considered in 
this article due to a phenomenon called non-collapsibility; see related discussion in  \cite{gail1984biased}, \cite{Freedman:2008ab} and \cite{fda:2019aa}. 
 Recently, \cite{wang2020modelrobust} has obtained some interesting results regarding  how to carry out  model-assisted inference on a linear contrast of $ \theta $ using logistic regression under stratified biased coin and stratified permuted block randomization.

Multiple treatment arms, which usually include a placebo, different doses (or regimens) of a new treatment,  and/or active controls,  are common in clinical trials \citep{Juszczak:2019aa} and are prevalent in some therapeutic areas such as immunology  \citep{Yates:2021aa}. In some applications,  the primary analysis may focus on comparing just two treatments,
even though the trial contains more than two treatment arms. 
A simple way of analysis is
to ignore the data from other arms and apply inference 
procedures to the two arms of interest. 
For ANOVA, this is  equivalent to using
all the arms, since ANOVA  does not borrow strength from other arms through using covariates. 
However, using data from all arms is
recommended for ANHECOVA, because it  utilizes  covariate data from  arms other than the two arms of interest to gain efficiency.
Regarding ANCOVA, there is no definite order of efficiency for using the whole dataset or data from two given arms, since using more covariate data in ANCOVA may increase or decrease efficiency.

Our \Cref{theo: CAR1} can also be  applied to rerandomization schemes
\citep{Li:2018aa, Li:2020aa} with discrete
covariates. Rerandomization  attempts to balance the treatment
assignments across levels of $Z$, but unlike sequential covariate-adaptive randomization, it randomizes the treatment assignments for all patients simultaneously. For two-armed trials, Corollaries 1 and 2 in 
\cite{Li:2018aa} show that rerandomization
satisfies (C2). Similar
results for model-assisted inference can also be found in
\citet[Theorem 3]{Li:2020aa}.

As a final cautionary note, standard software does not produce
asymptotically valid standard errors for model-assisted inference. We
implement an \textsf{R} package called \textsf{RobinCar} to compute the
model-assisted estimators and their robust standard errors, which is
available at the first author's website.

\vspace{-3.5mm}

\section*{Acknowledgements}
\vspace{-3mm}

The authors would like to thank the Associate Editor and two anonymous referees for helpful comments and suggestions.

\vspace{-3.5mm}

\section*{Supplementary Material}
\vspace{-3mm}

The supplementary material contains all technical proofs.

\spacingset{1.3}
\bibliographystyle{apalike}
\bibliography{reference}

\spacingset{1.5}

\begin{figure}[h]
	\centering
	\makebox{\includegraphics[scale=0.58]{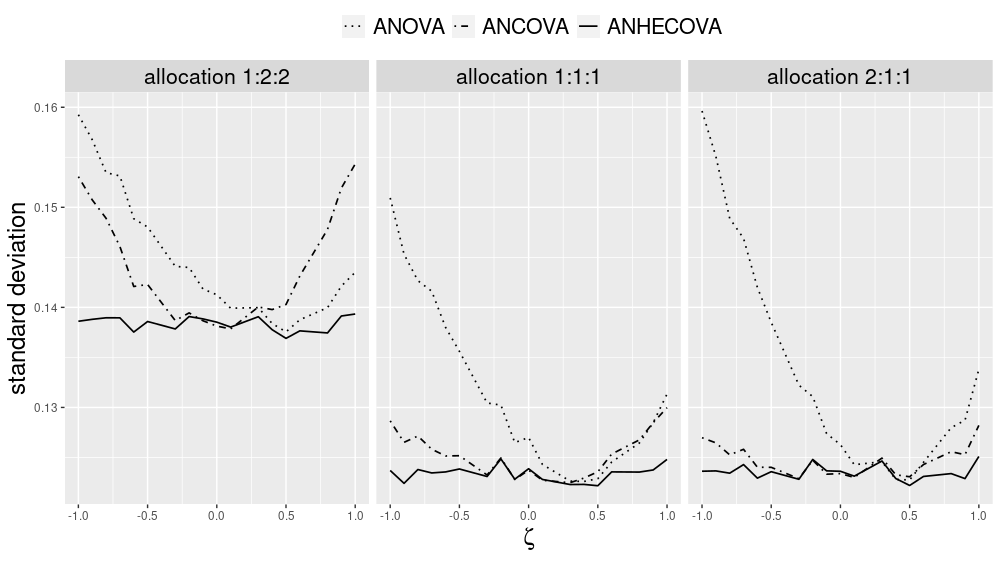}}
	\vspace{-6mm}
	\caption{\label{Fig:case1}{\small Standard deviations of ANOVA, ANCOVA, and ANHECOVA estimators based on 10,000 simulations and setup in \eqref{eq: simu1} }}
\end{figure}

\spacingset{1.5}
\begin{table}[h!]
	\begin{center}
		\caption{\small Bias, standard deviation (SD), average standard error (SE), and coverage probability (CP) of 95\% asymptotic confidence interval under simple randomization (SR),  stratified permuted block randomization (PB), and Pocock-Simon's minimization based on 10,000 simulations and setup in (\ref{eq: simu2}) with $ n=200 $ \label{tb:200}}
		\spacingset{1.0}
		\resizebox{\textwidth}{!}{
			\begin{tabular}{ccllccccccccc}
				\hline \\[-1.5ex]
				&               &           &          & \multicolumn{4}{c}{$\theta_2 - \theta_1 $} &  & \multicolumn{4}{c}{$\theta_3 - \theta_1 $} \\[0.5ex] \cline{5-8} \cline{10-13} \\[-1.5ex]
				Allocation &Randomization&  \multicolumn{1}{c}{Method} & \multicolumn{1}{c}{$X$}  & Bias      & SD       & SE       & CP       &  & Bias      & SD       & SE       & CP       \\[0.5ex] \hline \\[-1.5ex]
		1:1:1    &SR & ANOVA      &      & -0.002 & 0.467 & 0.463 & 0.944 &  & 0.000  & 0.284 & 0.285 & 0.950 \\
		&&ANCOVA& $ Z $      & 0.000  & 0.445 & 0.433 & 0.941 &  & -0.001 & 0.326 & 0.322 & 0.949 \\
		&&ANCOVA &$ Z,U,W  $  & -0.023 & 0.428 & 0.407 & 0.933 &  & 0.009  & 0.351 & 0.339 & 0.942 \\
	    &&ANHECOVA& $ Z $    & 0.000  & 0.384 & 0.372 & 0.939 &  & 0.003  & 0.238 & 0.235 & 0.943 \\
	 && ANHECOVA &$ Z,U,W  $& 0.000  & 0.325 & 0.315 & 0.943 &  & 0.001  & 0.220 & 0.213 & 0.941 \\
		&                  &        &       &       &       &  &        &       &       &       \\
		  & PB  & ANOVA   &         & -0.002 & 0.380 & 0.462 & 0.980 &  & -0.001 & 0.241 & 0.284 & 0.977 \\
		& &ANCOVA& $ Z $         & -0.002 & 0.379 & 0.432 & 0.972 &  & -0.001 & 0.242 & 0.321 & 0.991 \\
		& &ANCOVA &$ Z,U,W  $   & -0.026 & 0.356 & 0.406 & 0.970 &  & 0.009  & 0.275 & 0.338 & 0.983 \\
		& &ANHECOVA &$ Z $     & -0.002 & 0.377 & 0.371 & 0.940 &  & -0.001 & 0.240 & 0.234 & 0.940 \\
		& &ANHECOVA &$ Z,U,W  $ & -0.002 & 0.317 & 0.314 & 0.948 &  & -0.001 & 0.220 & 0.213 & 0.941 \\
		&                  &        &       &       &       &  &        &       &       &       \\
		 & Minimization & ANOVA  &          & 0.003  & 0.378 & 0.463 & 0.980 &  & 0.002  & 0.236 & 0.284 & 0.981 \\
		& &ANCOVA &$ Z $        & 0.003  & 0.378 & 0.432 & 0.972 &  & 0.002  & 0.237 & 0.321 & 0.991 \\
		 &&ANCOVA &$ Z,U,W  $   & -0.021 & 0.356 & 0.406 & 0.968 &  & 0.012  & 0.270 & 0.338 & 0.985 \\
		 &&ANHECOVA &$ Z $      & 0.002  & 0.376 & 0.372 & 0.946 &  & 0.002  & 0.236 & 0.234 & 0.947 \\
		& &ANHECOVA &$ Z,U,W  $ & 0.002  & 0.319 & 0.314 & 0.945 &  & 0.003  & 0.217 & 0.213 & 0.943  \\
			&          &         &        &       &       &       &  &        &       &       &       \\
		1:2:2&SR   & ANOVA    &         & 0.001  & 0.446 & 0.441 & 0.946 &  & 0.003  & 0.289 & 0.289 & 0.949 \\
		&& ANCOVA   & $ Z  $      & 0.004  & 0.430 & 0.417 & 0.942 &  & 0.002  & 0.347 & 0.339 & 0.946 \\
		&& ANCOVA   &$  Z, U, W $ & -0.019 & 0.420 & 0.398 & 0.934 &  & 0.012  & 0.380 & 0.365 & 0.943 \\
		&& ANHECOVA & $ Z  $      & 0.003  & 0.386 & 0.382 & 0.945 &  & 0.004  & 0.257 & 0.256 & 0.947 \\
		&& ANHECOVA & $ Z, U, W $ & 0.004  & 0.345 & 0.342 & 0.949 &  & 0.006  & 0.247 & 0.241 & 0.942 \\
		&          &         &        &       &       &       &  &        &       &       &       \\
		&PB   & ANOVA    &         & 0.002  & 0.379 & 0.441 & 0.977 &  & 0.000  & 0.254 & 0.289 & 0.971 \\
		&& ANCOVA   & $ Z $       & 0.002  & 0.378 & 0.414 & 0.968 &  & 0.000  & 0.257 & 0.337 & 0.988 \\
		&& ANCOVA   & $ Z, U, W $ & -0.024 & 0.365 & 0.395 & 0.961 &  & 0.008  & 0.296 & 0.362 & 0.982 \\
		&& ANHECOVA & $ Z  $      & 0.001  & 0.377 & 0.381 & 0.951 &  & 0.000  & 0.253 & 0.255 & 0.948 \\
		&& ANHECOVA & $ Z, U, W  $& 0.002  & 0.336 & 0.341 & 0.948 &  & 0.001  & 0.243 & 0.240 & 0.944 \\
		&&          &         &        &       &       &       &  &        &       &       &       \\
		 &Minimization& ANOVA    &         & 0.003  & 0.384 & 0.441 & 0.971 &  & 0.000  & 0.251 & 0.288 & 0.974 \\
		&& ANCOVA   & $ Z $       & 0.001  & 0.383 & 0.414 & 0.961 &  & -0.002 & 0.252 & 0.336 & 0.991 \\
		&& ANCOVA   & $ Z, U, W $ & -0.023 & 0.371 & 0.395 & 0.959 &  & 0.008  & 0.294 & 0.361 & 0.985 \\
		&& ANHECOVA &$  Z  $      & 0.002  & 0.382 & 0.381 & 0.944 &  & -0.001 & 0.250 & 0.254 & 0.950 \\
		&& ANHECOVA & $ Z, U, W $ & 0.000  & 0.338 & 0.341 & 0.952 &  & -0.001 & 0.239 & 0.239 & 0.948 \\[0.5ex] \hline
		\end{tabular}}
	\end{center}
\end{table}

\spacingset{1.5}

\begin{table}[h!]
	\begin{center}
		\caption{\small  Bias, standard deviation (SD), average standard error (SE), and coverage probability (CP) of 95\% asymptotic confidence interval under simple randomization (SR),  stratified permuted block randomization (PB), and Pocock-Simon's minimization based on 10,000 simulations and setup in (\ref{eq: simu2}) with $ n=400 $ \label{tb:400}}
		\spacingset{1.0}
		\resizebox{\textwidth}{!}{
			\begin{tabular}{ccllccccccccc}
				\hline \\[-1.5ex]
				&               &           &          & \multicolumn{4}{c}{$\theta_2 - \theta_1 $} &  & \multicolumn{4}{c}{$\theta_3 - \theta_1 $} \\[0.5ex] \cline{5-8} \cline{10-13} \\[-1.5ex]
				Allocation &Randomization&  \multicolumn{1}{c}{Method} & \multicolumn{1}{c}{$X$}  & Bias      & SD       & SE       & CP       &  & Bias      & SD       & SE       & CP       \\[0.5ex] \hline \\[-1.5ex]
			 1:1:1& SR           & ANOVA    &         & 0.001  & 0.327 & 0.327 & 0.946 &  & -0.001 & 0.201 & 0.201 & 0.949 \\
			&              & ANCOVA   & $ Z    $    & 0.001  & 0.310 & 0.306 & 0.945 &  & 0.000  & 0.229 & 0.227 & 0.949 \\
			&              & ANCOVA   & $ Z, U, W $ & -0.011 & 0.295 & 0.289 & 0.944 &  & 0.003  & 0.246 & 0.240 & 0.945 \\
			&              & ANHECOVA &$  Z  $      & -0.001 & 0.271 & 0.265 & 0.944 &  & -0.002 & 0.169 & 0.167 & 0.949 \\
			&              & ANHECOVA & $ Z, U, W $ & 0.000  & 0.226 & 0.224 & 0.947 &  & -0.002 & 0.155 & 0.152 & 0.948 \\
			&              &          &         &        &       &       &       &  &        &       &       &       \\
			& PB           & ANOVA    &         & 0.000  & 0.267 & 0.327 & 0.982 &  & -0.001 & 0.166 & 0.201 & 0.982 \\
			&              & ANCOVA   &$  Z  $      & 0.001  & 0.267 & 0.306 & 0.975 &  & -0.001 & 0.166 & 0.227 & 0.992 \\
			&              & ANCOVA   & $ Z, U, W $ & -0.012 & 0.250 & 0.289 & 0.974 &  & 0.003  & 0.186 & 0.240 & 0.987 \\
			&              & ANHECOVA & $ Z  $      & 0.001  & 0.266 & 0.265 & 0.947 &  & -0.001 & 0.165 & 0.167 & 0.950 \\
			&              & ANHECOVA &$  Z, U, W $ & 0.002  & 0.224 & 0.224 & 0.950 &  & -0.001 & 0.152 & 0.152 & 0.948 \\
			&              &          &         &        &       &       &       &  &        &       &       &       \\
			& Minimization & ANOVA    &         & -0.002 & 0.268 & 0.327 & 0.982 &  & 0.000  & 0.168 & 0.201 & 0.980 \\
			&              & ANCOVA   & $ Z  $      & -0.002 & 0.267 & 0.306 & 0.974 &  & 0.000  & 0.168 & 0.227 & 0.992 \\
			&              & ANCOVA   & $ Z, U, W $ & -0.013 & 0.250 & 0.289 & 0.974 &  & 0.005  & 0.189 & 0.240 & 0.987 \\
			&              & ANHECOVA & $ Z $       & -0.002 & 0.267 & 0.265 & 0.948 &  & 0.000  & 0.167 & 0.167 & 0.949 \\
			&              & ANHECOVA &$  Z, U, W $ & -0.001 & 0.225 & 0.223 & 0.946 &  & 0.000  & 0.154 & 0.152 & 0.944 \\ 
				&              &          &         &        &       &       &       &  &        &       &       &       \\
				 1:2:2& SR   & ANOVA    &         & 0.001  & 0.311 & 0.312 & 0.951 &  & 0.000 & 0.205 & 0.204 & 0.951 \\
				&      & ANCOVA   &$  Z  $      & 0.001  & 0.298 & 0.294 & 0.947 &  & 0.001 & 0.241 & 0.239 & 0.950 \\
				&      & ANCOVA   & $ Z, U, W $ & -0.013 & 0.286 & 0.281 & 0.945 &  & 0.003 & 0.262 & 0.257 & 0.947 \\
				&      & ANHECOVA & $ Z  $      & 0.000  & 0.268 & 0.268 & 0.949 &  & 0.001 & 0.182 & 0.180 & 0.948 \\
				&      & ANHECOVA & $ Z, U, W  $& 0.001  & 0.237 & 0.238 & 0.949 &  & 0.002 & 0.171 & 0.169 & 0.944 \\
				&      &          &         &        &       &       &       &  &       &       &       &       \\
				& PB   & ANOVA    &         & 0.002  & 0.265 & 0.312 & 0.977 &  & 0.001 & 0.180 & 0.204 & 0.975 \\
				&      & ANCOVA   & $ Z $       & 0.002  & 0.265 & 0.293 & 0.969 &  & 0.001 & 0.181 & 0.238 & 0.991 \\
				&      & ANCOVA   & $ Z, U, W $ & -0.012 & 0.253 & 0.280 & 0.967 &  & 0.005 & 0.207 & 0.256 & 0.986 \\
				&      & ANHECOVA & $ Z   $     & 0.002  & 0.264 & 0.267 & 0.950 &  & 0.001 & 0.179 & 0.179 & 0.949 \\
				&      & ANHECOVA &$  Z, U, W $ & 0.001  & 0.234 & 0.238 & 0.951 &  & 0.002 & 0.170 & 0.168 & 0.946 \\
				&      &          &         &        &       &       &       &  &       &       &       &       \\
				& Minimization & ANOVA    &         & 0.000  & 0.265 & 0.311 & 0.979 &  & 0.002 & 0.181 & 0.204 & 0.971 \\
				&      & ANCOVA   & $ Z   $     & -0.001 & 0.265 & 0.293 & 0.969 &  & 0.001 & 0.182 & 0.238 & 0.991 \\
				&      & ANCOVA   &$  Z, U, W  $& -0.013 & 0.253 & 0.280 & 0.970 &  & 0.006 & 0.208 & 0.256 & 0.985 \\
				&      & ANHECOVA & $ Z  $      & -0.001 & 0.265 & 0.267 & 0.951 &  & 0.001 & 0.181 & 0.179 & 0.947 \\
				&      & ANHECOVA & $ Z, U, W  $& 0.001  & 0.233 & 0.238 & 0.956 &  & 0.002 & 0.172 & 0.168 & 0.945\\[0.5ex] \hline
		\end{tabular}}
	\end{center}
\end{table}

\spacingset{1.5}

\begin{table}[h!]
  \begin{center}
    \caption{\small Estimate, standard error (SE), and p-value in the real data example analysis\label{tb3}}

    \spacingset{1.0}
    \resizebox{0.9\textwidth}{!}{
      \begin{tabular}{llcccccccc}
        \hline \\[-1.5ex]
        &         &  & \multicolumn{3}{c}{Physician versus placebo} &
        & \multicolumn{3}{c}{Soccer star versus placebo} \\[0.5ex] \cline{4-6} \cline{8-10} \\[-1.5ex]
        \multicolumn{1}{c}{Method} &   \multicolumn{1}{c}{$X$}    &  & Estimate      & SE         & p-value     &  & Estimate     & SE        & p-value    \\[0.5ex] \hline\\[-1.5ex]
        ANOVA &   &  & 0.386         & 0.211      & 0.067       &  & -0.068       & 0.205     & 0.739      \\
        \\[-1.5ex]
        ANCOVA          &       Grade         &  & 0.403         & 0.203      & 0.046       &  & -0.052       & 0.203     & 0.799      \\
        &                       Grade, Anemia status  &  & 0.437         & 0.199      & 0.028       &  & -0.085       & 0.201     & 0.672      \\
        \\[-1.5ex]
        ANHECOVA & Grade & & 0.409         & 0.200      & 0.041       &  & -0.051       & 0.201     & 0.800      \\
        & Grade, Anemia status &  & 0.481         & 0.193      & 0.013       &  & -0.046       & 0.195     & 0.815      \\[0.5ex] \hline
      \end{tabular}}
  \end{center}
\end{table}

\clearpage
\setcounter{footnote}{0}

  \begin{center}
	{\LARGE\bf Supplementary Material:  Toward Better Practice of Covariate Adjustment in Analyzing Randomized Clinical Trials} \\ \bigskip \bigskip
	{\large  Ting Ye\footnote{Department of Biostatistics, University of Washington.}, Jun Shao\footnote{School of Statistics, East China Normal University; Department of Statistics, University of Wisconsin.}, Yanyao Yi\footnote{Global Statistical Sciences, Eli Lilly and Company. } and Qingyuan Zhao\footnote{Department of Pure Mathematics and
			Mathematical Statistics, University of Cambridge.\\ Corresponding to Dr. Jun Shao. Email: {\tt shao@stat.wisc.edu}.}
	}
\end{center}


\setcounter{equation}{0}
\setcounter{table}{0}
\setcounter{lemma}{1}
\setcounter{section}{0}
\renewcommand{\theequation}{S\arabic{equation}}

\section{Two Lemmas}
\begin{lemma} \label{lem:identification}
	Assume (C1), (C2), and that $P(A_i=a_t\mid Z_1,\dots, Z_n)=\pi_t$ for all $ t=1,\dots, k $ and $ i=1,\dots,n$. We have the following conclusions. \\
	(i) For any integrable function $ f $,  
	$$
	E\{f(Y_i^{(t)}, X_i)\} = E(f(Y_i, X_i)\mid A_i = a_t)$$
	and
	$$
	E\{f(Y_i^{(t)}, X_i)\mid X_i \} = E(f(Y_i, X_i)\mid X_i, A_i = a_t ).
	$$
	(ii) Let $\theta = (E(Y^{(1)},...,E(Y^{(k)}))^{\top}$ be the potential response mean vector,
	${\beta} = \sum_{t=1}^k \pi_t \beta_t$, and $\beta_t =
	\Sigma_X^{-1} {\rm cov}( X_{i}, Y_i^{(t)})$, $t=1,...,k$. Then
	\[
	(\theta, {\beta}) =
	\arg\min_{(\vartheta,\gamma)} E\left[ \left\{Y_i - \vartheta^{\top} A_i  - \gamma^{\top}
	(X_i - \mu_X)\right\}^2\right]
	\]
	and
	\[
	(\theta, \beta_1, \dotsc,
	\beta_{k}) =
	\arg\min_{(\vartheta,\gamma_1,\dotsc, \gamma_k)}  E\left[ \left\{Y_i - \vartheta^{\top} A_i -
	\sum_{t=1}^k  \gamma_t^{\top} (X_i-\mu_X)I(A_i = a_t)\right\}^2\right].
	\]
\end{lemma}

The condition $ P(A_i=a_t\mid Z_1,\dots, Z_n)=\pi_t $ for all $ t $
and $ i $ holds for most covariate-adaptive randomization
schemes. Note that it
does not exclude the possibility that the set of
random variables $\{A_i,i=1,\dotsc,n\}$ is dependent of
$\{Z_i,i=1,\dotsc,n\}$, which is indeed the case for
covariate-adaptive randomization schemes. We impose this condition
only in Lemma 2 to facilitate understanding the working models. This additional
assumption is not needed for our asymptotic
theory in \S3, as condition (C2) is sufficient. 

\begin{proof}
	(i) We focus on proving the second result; the first  result can be shown
	similarly. For simple randomization, this result immediately follows  (C2)
	(i) as $(Y_i^{(1)}, \dotsc, Y_i^{(k)}, X_i, A_i)$ are independent and identically distributed. For
	covariate-adaptive randomization, we remark
	that the property of
	conditional independence \citep[Lemma 4.3]{Dawid:1979aa}, (C2) (i) and
	the third condition in Lemma \ref{lem:identification} imply that $A_i$ is independent of $\{(Y_i^{(1)},...,
	Y_i^{(k)}, X_i, Z_i), i=1,\dotsc,n\}$. Then, it can be shown that
	\begin{align*}
		E\{f(Y_i, X_i) \mid X_i, A_i = a_t\} 
		=& \ E\{f(Y_i^{(t)}, X_i) \mid X_i, A_i = a_t\} \\
		=& \sum_{z_1,\dotsc,z_n \in \mathcal{Z}} E\{f(Y_i^{(t)}, X_i)  \mid X_i, A_i =
		a_t, G_n\}
		P(G_n \mid X_i, A_i = a_t) \\
		=& \sum_{z_1,\dotsc,z_n \in \mathcal{Z}} E\{f(Y_i^{(t)}, X_i) \mid X_i, G_n\}
		P(G_n  \mid X_i, A_i = a_t) \\
		=& \sum_{z_1,\dotsc,z_n \in \mathcal{Z}} E\{f(Y_i^{(t)}, X_i)  \mid X_i, G_n\}
		P(G_n \mid X_i) \\
		=& \ E\{f(Y_i^{(t)}, X_i) \mid X_i\},
	\end{align*}
	where $G_n $ is the event that $\{Z_i = z_i, i=1,...,n\}$, and 
	the equalities follow from the consistency of
	potential responses, the law of iterated expectation, (C2) (i), and the remark above. \\
	(ii) We only prove the first result. The second result can be proved similarly.  Let $(\theta, \beta)$ be the optimality points satisfying the following estimation equations:
	\begin{align}
		&E\big[I(A_i=a_t) \{ Y_i - \theta^{\top} A_i - \beta^{\top} (X_i-\mu_X)\}\big]=0,\qquad \text{for any } t \label{eq: lem2-eq1}\\
		&E\big[(X_i- \mu_X) \{ Y_i - \theta^{\top} A_i - \beta^{\top} (X_i-\mu_X)\}\big]=0. \label{eq: lem2-eq2}
	\end{align}
	From Lemma \ref{lem:identification}(i), \eqref{eq: lem2-eq1} implies that for any $t$,  
	\begin{align*}
		E\big[ Y_i - \theta^{\top} A_i - \beta^{\top} (X_i-\mu_X)\mid A_i=a_t\big]=  E\big[ Y_i^{(t)} - \theta_t - \beta^{\top} (X_i-\mu_X)\big]=  E[ Y_i^{(t)} - \theta_t]=0 \label{eq: lem2-eq1}
	\end{align*}
	and, thus, $\theta_t= E(Y_i^{(t)})$, $t=1,...,k$. Then \eqref{eq: lem2-eq2} implies that 
	\begin{align*}
		0=&E\big[(X_i- \mu_X) \{ Y_i - \theta^{\top} A_i - \beta^{\top} (X_i-\mu_X)\}\big]\\
		=& \sum_{t=1}^k  E\big[I(A_i=a_t)(X_i- \mu_X) \{ Y_i - \theta^{\top} A_i - \beta^{\top} (X_i-\mu_X)\}\big]\\
		=& \sum_{t=1}^k  E\big[(X_i- \mu_X) \{ Y_i - \theta^{\top} A_i - \beta^{\top} (X_i-\mu_X)\}\mid A_i=a_t\big]\pi_t\\
		=& \sum_{t=1}^k  E\big[(X_i- \mu_X) \{ Y^{(t)}_i - \theta_t - \beta^{\top} (X_i-\mu_X)\}\big]\pi_t\\
		=& \sum_{t=1}^k \left[ {\rm cov}(X_i, Y_i^{(t)})- \Sigma_X \beta \right] \pi_t\\
		=&  \sum_{t=1}^k  {\rm cov}(X_i, Y_i^{(t)})\pi_t - \Sigma_X \beta  
	\end{align*}
	and, thus, $ \beta= \Sigma_X^{-1}  \sum_{t=1}^k  {\rm cov}(X_i, Y_i^{(t)})\pi_t= \sum_{t=1}^k \pi_t \beta_t$. 
\end{proof}

\begin{lemma} \label{lemma: coef}
	Under conditions (C1)-(C2),	for $ t=1,\dots, k $, $\hat{\beta}_t= \beta_t+o_p(1)$ 
	and $\hat{\beta} = \beta+o_p(1)$; 
\end{lemma}

\noindent
\begin{proof}
	(i) We prove the result for $\hat{\beta}_t$.
	The proof for $\hat{\beta}$ is analogous and  omitted. Notice that
	\[
	\frac{1}{n_t}  \sum_{ i : A_i = a_t} (X_i -\bar X_t) Y_i =  \frac{1}{n_t}  \sum_{i=1}^n  I(A_i=a_t) X_i Y_i-\frac{1}{n_t} \sum_{i=1}^n I(A_i=a_t) X_i  \frac{1}{n_t}   \sum_{i=1}^n  I(A_i=a_t)  Y_i 
	\]
	Let $\mathcal{A}=\{A_1, \dots, A_n\}$ and  $\mathcal{F}=\{Z_1, \dots, Z_n\}$.  Note that  
	\begin{align*}
		E \left\{   \frac{1}{n}   \sum_{i=1}^n  I(A_i=a_t) X_i Y_i\mid \mathcal{A}, \mathcal{F}  \right\} & =  \frac{1}{n}  \sum_{i=1}^n I (A_i=a_t) E(X_i Y_i^{(t)} \mid \mathcal{A}, \mathcal{F} ) \\
		&= \frac{1}{n}   \sum_{i=1}^n I (A_i=a_t) E(X_i Y_i^{(t)}  \mid Z_i),
	\end{align*}
	where the second line holds because $ E(X_i Y_i^{(t)}  \mid \mathcal{A}, \mathcal{F}) = E(X_i Y_i^{(t)}  \mid \mathcal{F}) =  E(X_i Y_i^{(t)}  \mid Z_i)$ from (C1) and (C2) (i). Moreover, $  n^{-1}\sum_{i=1}^n  I(A_i=a_t) X_i Y_i^{(t)} $ is an average of independent  random variables once conditional on $\{\mathcal{A}, \mathcal{F} \}$. From the existence of second moment of $X Y^{(t)}$, and the weak law of large numbers for independent random variables, we conclude that, for any $\epsilon>0$,
	\[
	\lim_{n\rightarrow \infty} P\left( \bigg| \frac{1}{n}   \sum_{i=1}^n I(A_i=a_t) X_i Y_i-  \frac{1}{n}   \sum_{i=1}^n I (A_i=a_t) E(X_i Y_i^{(t)}  \mid Z_i) \bigg|   \geq  \epsilon \, \mid  \, \mathcal{A}, \mathcal{F}\right)=0
	\]
	From the bounded convergence theorem, the above equation also holds unconditionally. In other words, 
	\begin{align*}
		\frac{1}{n}   \sum_{i=1}^n I(A_i=a_t) X_i Y_i-  \frac{1}{n}   \sum_{i=1}^n I (A_i=a_t) E(X_i Y_i^{(t)}  \mid Z_i) = o_p(1).
	\end{align*}
	Furthermore, 
	\begin{align*}
		\frac{1}{n}   \sum_{i=1}^n I (A_i=a_t) E(X_i Y_i^{(t)}  \mid Z_i) 
		&= \frac{1}{n} \sum_{z}  \sum_{i=1}^n I(Z_i=z) I (A_i=a_t) E(X_i Y_i^{(t)}  \mid Z_i=z)\\
		&= \frac{1}{n} \sum_{z} E(X_i Y_i^{(t)}  \mid Z_i=z) \sum_{i=1}^n I(Z_i=z) I (A_i=a_t) \\
		&= \frac{1}{n} \sum_{z} E(X_i Y_i^{(t)}  \mid Z_i=z) n_t(z) \\
		&=  \sum_{z} E(X_i Y_i^{(t)}  \mid Z_i=z)\frac{ n_t(z)}{n(z)}  \frac{n(z)}{n}\\
		&= \sum_{z} E(X_i Y_i^{(t)}  \mid Z_i=z)\pi_t P(Z_i=z)  + o_p(1)\\
		&= \pi_t E( X_iY_i^{(t)} )  + o_p(1) 
	\end{align*}
	This together with the fact that  $ n_t/n= \sum_{z} n_t(z) / \{ \sum_{z} n(z) \}= \pi_t +o_p(1)$,  we have 
	\begin{align*}
		\frac{1}{n_t}   \sum_{i=1}^n I(A_i=a_t) X_i Y_i= E( X_iY_i^{(t)})  + o_p(1) 
	\end{align*}
	Similarly, we can show the result with $X_iY_i$ replaced by $X_i$ or $Y_i$ also holds, i.e., 
	\begin{align*}
		&\frac{1}{n_t}   \sum_{i=1}^n I(A_i=a_t) X_i= E( X_i )  + o_p(1) \\
		&\frac{1}{n_t}   \sum_{i=1}^n I(A_i=a_t) Y_i= E( Y_i^{(t)} )  + o_p(1) 
	\end{align*}
	The denominator of $\hat{\beta}_t$ can be treated similarly, which leads to
	$$
	\frac{1}{n_t} \sum_{ i: A_i=a_t } \{X_i - \bar{X}_t\}\{X_i - \bar{X}_t\}^{\top} =\Sigma_X+o_p(1).
	$$
	The proof is completed by using the definition of $ \beta_t$.

\end{proof}  

\section{Technical Proofs}
\subsection{Proof of (9)}
Under simple randomization,  $ A_1, \dots, A_n$ are independent with other variables.  
Let  $ \bar{X}_{-t} = (n-n_t)^{-1} \sum_{i: A_i\neq a_t} X_i  $. Then 
$		\bar X_t - \bar X  =  n^{-1}(n- n_t) ( \bar X_t- \bar X_{-t}) $. 
Note  that $ \bar X_t$ and  $\bar X_{-t} $ are uncorrelated conditional on $ \mathcal{A} = (A_1,...,A_n)$, as
\begin{align*}
	{\rm cov} (\bar X_t, \bar X_{-t}\mid  \mathcal{A})  = \frac{1}{(n-n_t) n_t} \sum_{i=1}^n\sum_{j=1}^n I(A_i=a_t)  I(A_j\neq a_t) {\rm cov} (X_i, X_j \mid \mathcal{A}) =0,
\end{align*}
where the last equality is from $ {\rm cov} (X_i, X_j \mid\mathcal{A})= {\rm cov} (X_i, X_j  ) = 0$ for $ i\neq j $. Similarly, we can show that $ \bar Y_t $ and  $\bar X_{-t} $ are uncorrelated conditional on $ \mathcal{A}$.

Then, 
\begin{align*}
	{\rm cov}\{  \beta_t^{\top} (\bar X_t - \bar X), \bar Y_t\}  
	&= \beta_t^{\top}  {\rm cov}\left(   \frac{n- n_t}{n} \bar X_t- \frac{n-n_t}{n} \bar X_{-t}, \bar Y_t \right)  \\
	&=  \beta_t^{\top} E \left\{ {\rm cov}\left(   \frac{n- n_t}{n} \bar X_t, \bar Y_t\mid {\cal A}  \right) \right\}\\
	&=  \beta_t^{\top} E\left\{  \frac{n- n_t}{nn_t^2}  {\rm cov} \left( \sum_{ i : A_i = a_t} X_i  , \sum_{ i : A_i = a_t} Y_i \mid \mathcal{A}\right) \right\}   \\
	&=  \beta_t^{\top} E\left\{  \frac{n- n_t}{nn_t^2}\sum_{ i : A_i = a_t}  {\rm cov} \left(  X_i  ,  Y_i^{(t)} \right) \right\}    \\
	&=  \beta_t^{\top} E\left\{  \frac{n- n_t}{nn_t} \right\}   {\rm cov} \left(  X_i  ,  Y_i^{(t)} \right)\\
	&=E\left\{  \frac{n- n_t}{nn_t} \right\}   \beta_t^{\top}   \Sigma_X \beta_t
\end{align*}
where the second equality is from $ 	{\rm cov} (\bar X_{-t}, \bar Y_t\mid  \mathcal{A})   =0$, $ E(\bar Y_t\mid \mathcal{A}) = E(Y^{(t)})$ and  the identity that $ {\rm cov} (X, Y)= E\{ {\rm cov} (X, Y\mid Z)\}  + {\rm cov}\{ E(X\mid Z), E(Y\mid Z)\}$. 
Also note that 
\begin{align*}
	{\rm var} \{ \beta_t^{\top} (\bar X_t- \bar X)\} &=\beta_t^{\top} {\rm var}\left(   \frac{n- n_t}{n} ( \bar X_t- \bar X_{-t})   \right)\beta_t   \\
	&= \beta_t^{\top} E\left(   \frac{(n- n_t)^2}{n^2} {\rm var} ( \bar X_t- \bar X_{-t}\mid \mathcal{A} )  \right)\beta_t   \\
	&= \beta_t^{\top} E\left(   \frac{(n- n_t)^2}{n^2}\{ {\rm var} ( \bar X_t\mid \mathcal{A} ) + {\rm var} ( \bar X_{-t}\mid  \mathcal{A} ) \}\right)\beta_t   \\ 
	&= \beta_t^{\top} E\left(   \frac{(n- n_t)^2}{n^2}\left\{\frac{{\var (X_i)}}{n_t}+\frac{{\var (X_i)}}{n- n_t} \right\}\right)\beta_t   \\ 
	&= E\left\{  \frac{n- n_t}{nn_t} \right\}   \beta_t^{\top}   \Sigma_X \beta_t
\end{align*}
where the second equality uses the identity that $ {\rm var} (X)= E \{ {\rm var} (X\mid Z) \}+   {\rm var} \{E(X\mid Z) \}$, and $ E\left(  \bar X_t- \bar X_{-t}\mid \mathcal{A} \right) = E(X_i)- E(X_i)=0 $.

\subsection{Proof of Lemma 1}
For any fixed $k$-dimensional vector $\ell= (\ell_1,\dots, \ell_k)^\top$, we have
\begin{eqnarray*}
	&&\ell^\top \{{\rm diag}(\pi_t^{-1} m_t^\top m_t) - M^\top M\}\ell \\
	&=&  \sum_{t=1}^k \pi_t^{-1}\ell_t^2m_t^\top m_t -  \left\{\sum_{t=1}^k \ell_t m_t^\top\right\}\left\{\sum_{t=1}^k \ell_t m_t\right\}\\
	&=&E(Q^\top Q)-E(Q^\top)E(Q) \\
	&=&  \text{tr}\{E(Q Q^{\top})\}- \text{tr}\{E(Q) E(Q^{\top})\}\\
	&\geq& 0,
\end{eqnarray*}
where $ {\rm tr} $ denotes the trace of a matrix, $ Q $ denotes a $
p $-dimensional random vector that  takes value $ \pi_t^{-1}\ell_t
m_t$ with probability $\pi_t$, $ t=1,
\dots, k $, and the last equality follows from the fact that the covariance matrix $
\text{var} (Q) = E(Q Q^{\top})- E(Q) E(Q^{\top})$ is positive semidefinite.

\subsection{Proof of Theorem 1}
(i) First, from $\bar X_t- \bar X= O_p(n^{-1/2})$ and $\hat b_t= b_t+o_p(1)$, we have
\begin{eqnarray*}
	\hat\theta(\hat b_1, \dots, \hat b_k)&=&  \hat\theta( b_1, \dots,  b_k)+ \{ (\bar X_1- \bar X) (b_1- \hat b_1), \dots,(\bar X_1- \bar X) (b_k- \hat b_k) \}^{\top} \\
	&=& \hat\theta( b_1, \dots,  b_k)+o_p(n^{-1/2}). 
\end{eqnarray*}
Write the sample average as  $ \mathbb{E}_n [\mu(X) ]= n^{-1} \sum_{i=1}^{n} \mu(X_i) $. Then, 
\begin{align*}
	\bar X - \mu_X &= \sum_{t=1}^{k} \frac1n \sum_{i=1}^n I (A_i= a_t) (X_i - \mu_X) = \sum_{t=1}^{k}\mathbb{E}_n \left[ I(A=a_t) (X- \mu_X)  \right] ,
\end{align*}
and 
\begin{align*}
	&  \bar Y_t- \theta_t- (\bar X_t- \mu_X)^{\top} b_t   \\
	&=  \frac{1}{n_t} \sum_{i=1}^n I(A_i=a_t)  \left\{ 
	Y_i - \theta_t - (X_i -\mu_X)^\top b_t\right\} \\
	&= \pi_t^{-1}   \frac{1}{n}  \sum_{i=1}^n I(A_i=a_t)  \left\{ 
	Y_i - \theta_t - (X_i -\mu_X)^\top b_t\right\} \\
	&\qquad +  \left( \frac{1}{n_t/n} -   \frac{1}{\pi_t}    \right)  \frac{1}{n}  \sum_{i=1}^n I(A_i=a_t)  \left\{ 
	Y_i - \theta_t - (X_i -\mu_X)^\top b_t\right\} \\
	&= \pi_t^{-1} \mathbb{E}_n \left[ I(A=a_t) \{ Y- \theta_t- (X-\mu_X)^{\top} b_t\} \right]   + o_p(n^{-1/2}),
\end{align*} 
where the last equality holds because $  \mathbb{E}_n \left[ I(A=a_t) \{ Y- \theta_t- (X-\mu_X)^{\top} b_t\} \right]   = O_p(n^{-1/2}) $ from the central limit theorem, and  $  n/n_t - \pi_t^{-1} = o_p(1)$ from condition (C2) (ii).  Hence, we can decompose $\hat\theta( b_1, \dots,  b_k)$ as
\begin{align*}
	&\hat\theta( b_1, \dots,  b_k)- \theta \\
	&= \left(
	\begin{array}{c}
		\bar Y_1- \theta_1- (\bar X_1- \mu_X)^{\top} b_1 \\
		\vdots\\
		\bar Y_k- \theta_k- (\bar X_k- \mu_X)^{\top} b_k
	\end{array}\right) +
	\left(
	\begin{array}{c}
		b_1^\top  (\bar X - \mu_X )    \\
		\vdots \\
		b_k^\top  (\bar X - \mu_X ) 
	\end{array}\right)  \\
	&  =\underbrace{\left(
		\begin{array}{c}
			\pi_1^{-1} \mathbb{E}_n \left[ I(A=a_1) \{ Y- \theta_1- (X-\mu_X)^{\top} b_1\} \right]    \\
			\vdots\\
			\pi_k^{-1} \mathbb{E}_n \left[ I(A=a_k) \{ Y- \theta_k- (X-\mu_X)^{\top} b_k\} \right]   
		\end{array}\right)}_{M_1} +
	\underbrace{ \left(
		\begin{array}{c}
			b_1^\top \sum_{t=1}^{k}\mathbb{E}_n \left[ I(A=a_t) (X- \mu_X)  \right]     \\
			\vdots\\
			b_k^\top  \sum_{t=1}^{k}\mathbb{E}_n \left[ I(A=a_t) (X- \mu_X)  \right]  
		\end{array}\right)}_{M_2} \\
	&\qquad + o_p(n^{-1/2})   \\   
	& =  \left(
	\begin{array}{ccccc}
		\pi_1^{-1}  a_1^{\top}  &  -	\pi_1^{-1}  b_1^{\top}  & {0}_p^{\top}  &\cdots & {0}_p^{\top}     \\
		\vdots &	\vdots & \vdots & \ddots  & \vdots\\
		\pi_k^{-1}    a_k^{\top}  &  {0}_p^{\top}   &  {0}_p^{\top}  &\cdots & - 	\pi_k^{-1}  b_k^{\top} 
	\end{array}\right)_{k\times (k+kp)}
	\underbrace{\left(
		\begin{array}{c}
			\mathbb{E}_n \left[ I(A=a_1) ( Y- \theta_1) \right]   \\
			\vdots\\
			\mathbb{E}_n \left[ I(A=a_k) ( Y- \theta_k) \right]   \\ 
			\mathbb{E}_n \left[ I(A=a_1) (X- \mu_X)  \right]   \\
			\vdots\\
			\mathbb{E}_n \left[ I(A=a_k) (X- \mu_X)  \right]  
		\end{array}\right)}_{V_{(k+kp)\times 1}} \\
	&   +  
	\left(
	\begin{array}{ccccc}
		b_1^\top  &  b_1^\top   &\dots &  b_1^\top   \\
		\vdots &\vdots &\ddots &\vdots \\
		b_k^\top   &b_k^\top   &\dots &    b_k^\top  
	\end{array}\right)_{k\times (kp) }
	\left(
	\begin{array}{c}
		\mathbb{E}_n \left[ I(A=a_1) (X- \mu_X)  \right]   \\
		\vdots\\
		\mathbb{E}_n \left[ I(A=a_k) (X- \mu_X)  \right]   
	\end{array}\right)_{(k+kp)\times 1}  + o_p(n^{-1/2}) ,
\end{align*}
where $ a_t $ denotes the $ k $-dimensional vector whose $ t $th component is 1 and other components are 0, $ p $ is the dimension of $ X $,  $ 0_p $ denotes a $ p $-dimensional vector of zeros. From the central limit theorem, we have that the random vector $\sqrt{n} V $ is asymptotically normal with mean 0. This implies that  $\sqrt{n} \{\hat\theta( b_1, \dots,  b_k)- \theta\}$ is asymptotically normal with mean 0 from the Cram\'{e}r-Wold device.

It remains to calculate the asymptotic variance of $\sqrt{n} \{\hat\theta( b_1, \dots,  b_k)- \theta\}$. In the following, we consider $ M_1 $ and $ M_2 $ separately. 

Consider $M_1$, where the $t$th component equals
\[
M_{1t}  =  	\pi_t^{-1} \mathbb{E}_n \left[ I(A=a_t) \{ Y- \theta_t- (X-\mu_X)^{\top} b_t \} \right]   . 
\]
We have that $( M_{1t}, t=1,\dots, k) $ are mutually independent and 
\[
{\rm var}(M_{1t} )= (n\pi_t)^{-1}{\rm var} \{Y^{(t)}- X^{\top} b_t\}, 
\]
Hence, ${\rm var} (M_1)$ is a diagonal matrix, with the diagonal elements being ${\rm var}(M_{1t} ), t=1,\dots, k$. 
That is,  $ n{\rm var} (M_1  )= {\rm diag} \{ \pi_t^{-1}{\rm var} (Y^{(t)}- X^{\top} b_t)\} $.

Next, consider $M_2$, which can be reformulated as
\begin{displaymath}
	M_2=\left(
	\begin{array}{ccc}
		b_1^{\top}&& \\
		&\ddots&\\
		&&b_k^{\top}
	\end{array}\right)_{k\times (kp)}
	\left(
	\begin{array}{c}
		\sum_{t=1}^{k}\mathbb{E}_n \left[ I(A=a_t) (X- \mu_X)  \right]    \\
		\cdots\\
		\sum_{t=1}^{k}\mathbb{E}_n \left[ I(A=a_t) (X- \mu_X)  \right]   
	\end{array}\right)_{ (kp)\times 1}
\end{displaymath}
whose variance  can be easily derived as $ n {\rm var}  (M_2)  = B^{\top} \Sigma_X B $. 

Finally, consider ${\rm cov} (M_1,  M_2)$ whose $(t, s)$ element equals
\begin{eqnarray*}
	&&{\rm cov} \left\{   	\pi_t^{-1} \mathbb{E}_n \left[ I(A=a_t) \{ Y- \theta_t- (X-\mu_X)^{\top} b_t
	\} \right], 		b_s^\top \sum_{t=1}^{k}\mathbb{E}_n \left[ I(A=a_t) (X- \mu_X)  \right]   \right\}\\
	&&= {\rm cov} \left\{   	\pi_t^{-1} \mathbb{E}_n \left[ I(A=a_t) \{ Y- \theta_t- (X-\mu_X)^{\top} b_t
	\} \right], 		b_s^\top  \mathbb{E}_n \left[ I(A=a_t) (X- \mu_X)  \right]   \right\}\\
	&&=n^{-1}  \pi_t^{-1}  {\rm cov} \left\{    I(A=a_t)  ( Y- X^{\top} b_t) 
	\} , 		b_s^\top    I(A=a_t) (X- \mu_X)   \right\}\\
	&& = n^{-1}  \pi_t^{-1}  E \left\{    I(A=a_t)  ( Y- X^{\top} b_t) 
	b_s^\top    (X- \mu_X)   \right\}\\ 
	&& = n^{-1}    E \left\{    ( Y^{(t)}- X^{\top} b_t) 
	b_s^\top    (X- \mu_X)   \right\}\\ 		
	&& = n^{-1}   \left\{   {\rm cov}     ( Y^{(t)} ,
	b_s^\top    X )   -  {\rm cov}  (    X^\top b_t, 
	b_s^\top    X )  \right\}    \\ 		
	&&= n^{-1}  \{ \beta_t^\top \Sigma_X b_s  - b_t^\top \Sigma_X b_s \} \\
	&&= n^{-1}  ( \beta_t - b_t )^\top \Sigma_X b_s .
\end{eqnarray*}
Thus, $ n{\rm cov} (M_1, M_2)= (\mathscr{B}- B)^\top \Sigma_X B $ and $ n{\rm cov} (M_2, M_1)= B^\top \Sigma_X   (\mathscr{B}- B) $. 
Combining the above results, we conclude that $\sqrt{n} \{\hat\theta( b_1, \dots,  b_k)- \theta\}$ is asymptotically normal with mean 0 and variance $V_{\rm SR}(B)$, 
\begin{align*}
	V_{\rm SR}(B)&={\rm diag} \{\pi_t^{-1}{\rm var} (Y^{(t)}- X^\top b_t)\}+ (\mathscr{B}- B)^\top \Sigma_X B+ B^\top \Sigma_X   (\mathscr{B}- B)  + B^{\top} \Sigma_X B \\
	&= {\rm diag} \{\pi_t^{-1}{\rm var} (Y^{(t)}- X^\top b_t)\}+ \mathscr{B}^\top \Sigma_X B+ B^\top \Sigma_X   \mathscr{B}  - B^{\top} \Sigma_X B  .
\end{align*}
(ii) Note that 
\begin{align*}
	{\rm var} (Y^{(t)}- b_t^{\top} X)  =& {\rm var} (Y^{(t)}- \beta_t^{\top}  X+\beta_t^{\top}  X- b_t^{\top}X)  \\
	=&   {\rm var} \{Y^{(t)}-\beta_t^{\top} X\}+{\rm var} \{  (\beta_t - b_t)^{\top} X\} + 2{\rm cov} \{ Y^{(t)}- \beta_t^{\top}X,(\beta_t - b_t)^{\top} X \} \\
	=& {\rm var} \{Y^{(t)}- \beta_t^{\top} X\}+ (\beta_t - b_t)^{\top} \Sigma_X (\beta_t - b_t)  .
\end{align*}
Then simple algebra shows that 
\begin{align*}
	&	V_{\rm SR} (B) - 	V_{\rm SR} (\mathscr{B})  \\
	=	& \ {\rm diag} \{ \pi_t^{-1} {\rm var} (Y^{(t)}- b_t^{\top} X)\} -  {\rm diag} \{ \pi_t^{-1} {\rm var} (Y^{(t)}- \beta_t^{\top} X)\}  -  (\mathscr{B}- B)^{\top} \Sigma_X  (\mathscr{B}- B) \\
	= & \ {\rm diag} \{ \pi_t^{-1}(\beta_t - b_t)^{\top} \Sigma_X (\beta_t - b_t)   \} -  (\mathscr{B}- B)^{\top} \Sigma_X  (\mathscr{B}- B).
\end{align*}
The rest follows from applying Lemma 1 with $ M= \Sigma_X^{1/2} (\mathscr{B}- B) $.

\subsection{Proof of Corollary 1}
From Lemma \ref{lemma: coef}, we know that $ \hat\beta= \beta+o_p(1) $ and $ \hat\beta_t= \beta_t+o_p(1) $, $ t=1,\dots, k $. Let $ \sigma_A^2, \sigma_B^2, \sigma_U^2 $ respectively be the asymptotic variance of $ \sqrt{n}c_{ts}^{\top}\hancova$, $ \sqrt{n}c_{ts}^{\top} \ancova$ and $   \sqrt{n} c_{ts}^{\top}\anova$, where from Theorem 1,
\begin{align*}
	\sigma_A^2 = & \frac{\var (Y^{(t)}- X^{\top} \! \beta_t)}{\pi_t} + \frac{\var (Y^{(s)}- X^{\top} \! \beta_s)}{\pi_s}  + (\beta_t- \beta_s)^{\top}  \Sigma_X(\beta_t- \beta_s)
	\\
	\sigma^2_B =&
	\frac{\var ( Y^{(t)}- X^{\top} \! \beta) }{\pi_t} +
	\frac{\var ( Y^{(s)}- X^{\top} \! \beta ) }{\pi_s}\\
	\sigma_U^2 = & \frac{\var (Y^{(t)})}{\pi_t} + \frac{\var (Y^{(s)})}{\pi_s}
\end{align*}
The results in Corollary 1(i) follows from
\begin{align*}
	&\sigma_A^2 - \sigma_U^2 \nonumber\\
	= & \, \frac{\beta_t^{\top} \Sigma_X \beta_t -2{\rm cov}(  X^{\top} \! \beta_t, Y^{(t)}) }{\pi_t}+\frac{ \beta_s^{\top} \Sigma_X \beta_s -2{\rm cov}(  X^{\top} \! \beta_s, Y^{(s)})}{\pi_s}+ \{ \beta_t- \beta_s\}^{\top} \Sigma_X\{\beta_t-\beta_s\} \vspace{2mm}\nonumber \\
	= & \, \frac{ \beta_t^{\top} \Sigma_X \beta_t -2 \beta_t^{\top} \Sigma_X \beta_t}{\pi_t} + \frac{\beta_s^{\top} \Sigma_X \beta_s - 2 \beta_s^{\top} \Sigma_X \beta_s  }{\pi_s}   + \{ \beta_t- \beta_s\}^{\top} \Sigma_X\{\beta_t-\beta_s\} \vspace{2mm}\nonumber \\
	=&\,  - \frac{ \beta_t^{\top} \Sigma_X \beta_t }{\pi_t} -\frac{ \beta_s^{\top} \Sigma_X \beta_s }{\pi_s}+\{ \beta_t- \beta_s\}^{\top} \Sigma_X\{\beta_t-\beta_s\}    \vspace{2mm}\\
	= &\,   -\frac{\{\pi_s \beta_t+\pi_t\beta_s\}^{\top} \Sigma_X\{\pi_s \beta_t+\pi_t\beta_s\}}{\pi_t\pi_s (\pi_t+\pi_s)}-\{ \beta_t- \beta_s\}^{\top} \Sigma_X\{\beta_t-\beta_s\} \left(\frac{1-\pi_t-\pi_s}{\pi_t+\pi_s}\right)\nonumber
\end{align*}
where the second equality follows  from $\beta_t = \Sigma_X^{-1} {\rm cov} (X , Y^{(t)}  )$.
This also proves that $\sigma_A^2 \leq \sigma_U^2$, because
$\Sigma_X $ is positive definite  and $ \pi_t+\pi_s\leq 1 $. If $\sigma_A^2 = \sigma_U^2$, then we must have $\pi_s \beta_t+\pi_t\beta_s =0$ and $(1-\pi_t-\pi_s) \{\beta_t- \beta_s\}=0  $.

To show the results  in Corollary 1(ii), notice that
\begin{align*}
	\sigma^2_B
	=& \frac{\var \{Y^{(t)}- X^{\top} \! \beta_t+ X^{\top} \! \beta_t- X^{\top} \! \beta \}}{\pi_t} +  \frac{\var \{Y^{(s)}- X^{\top} \! \beta_s+ X^{\top} \! \beta_s- X^{\top} \! \beta \}}{\pi_s} \\
	=&\frac{\var \{Y^{(t)}- X^{\top} \! \beta_t\}+ \var\{X^{\top} \! \beta_t- X^{\top} \! \beta\}}{\pi_t} + \frac{\var \{Y^{(s)}- X^{\top} \! \beta_s \}+\var\{ X^{\top} \! \beta_s- X^{\top} \! \beta \}}{\pi_s}
\end{align*}
where the second equality holds because
\begin{align*}
	& {\rm cov} \{ Y^{(t)}- \beta_t^{\top} X, \beta_t^{\top} X- \beta^{\top} X \}  = {\rm cov} \{ Y^{(t)}-\beta_t^{\top} X, X  \} \{\beta_t-\beta \} \\
	&= \{ {\rm cov} ( Y^{(t)}, X    )- \beta_t^{\top} \Sigma_X\} \{\beta_t-\beta \} =0
\end{align*}
Then,
\begin{align*}
	&\sigma_A^2- \sigma_B^2=  \{\beta_t- \beta_s\}^{\top}  \Sigma_X\{\beta_t- \beta_s\}-\frac{ \{ \beta_t- \beta\}^{\top}\Sigma_X\{ \beta_t- \beta\}}{\pi_t}- \frac{ \{ \beta_s- \beta\}^{\top}\Sigma\{ \beta_s- \beta\}}{\pi_s}
\end{align*}
In order to show that $ \sigma_A^2-\sigma_B^2\leq 0  $, we prove a stronger statement:   it is true that for any $ \tilde\beta $,
\begin{align}
	&  \{\beta_t- \beta_s\}^{\top}  \Sigma_X\{\beta_t- \beta_s\}-\frac{ \{ \beta_t-\tilde \beta\}^{\top}\Sigma_X\{ \beta_t- \tilde\beta\}}{\pi_t}- \frac{ \{ \beta_s- \tilde\beta\}^{\top}\Sigma_X\{ \beta_s-\tilde \beta\}}{\pi_s}\leq 0. \label{eq: state}
\end{align}
As a consequence, setting $ \tilde\beta $ as $ \beta = \sum_{t=1}^k\pi_t\beta_t$, the statement in (\ref{eq: state}) also holds. This proves $ \sigma_A^2- \sigma_B^2\leq 0 $.

In what follows, we prove the claim in (\ref{eq: state}). Note that the gradient of the left hand side of (\ref{eq: state}) is
\begin{align*}
	- 2\left[ \frac{\{\tilde\beta- \beta_t \}^{\top} \Sigma_X }{\pi_t}+\frac{\{\tilde\beta- \beta_s \}^{\top} \Sigma_X  }{\pi_s} \right],
\end{align*}
which equals zero when $ \tilde\beta= \{ \pi_s\beta_t+ \pi_t\beta_s\}/(\pi_t+\pi_s) $. This is also the unique solution from the positive definiteness of $ \Sigma_X  $. It is also easy to see that the Hessian of the left hand side of (\ref{eq: state}) is negative definite, which means that $ \tilde\beta= \{ \pi_s\beta_t+ \pi_t\beta_s\}/(\pi_t+\pi_s) $ is the global and unique maximizer of the left hand side of (\ref{eq: state}). The statement in (\ref{eq: state}) is true because when evaluated at $ \tilde\beta= \{ \pi_s\beta_t+ \pi_t\beta_s\}/(\pi_t+\pi_s) $, the left hand side of (\ref{eq: state}) equals
\begin{align*}
	&  \{\beta_t- \beta_s\}^{\top}  \Sigma_X \{\beta_t- \beta_s\}\nonumber\\
	&- \left\{\beta_t- \frac{\pi_s\beta_t + \pi_t\beta_s}{\pi_t+\pi_s}\right\}^{\top} \Sigma_X   \left\{\beta_t- \frac{\pi_s\beta_t + \pi_t\beta_s}{\pi_t+\pi_s}\right\}\frac{1}{\pi_t}\\
	&- \left\{\beta_s- \frac{\pi_s\beta_t + \pi_t\beta_s}{\pi_t+\pi_s}\right\}^{\top} \Sigma_X   \left\{\beta_s- \frac{\pi_s\beta_t + \pi_t\beta_s}{\pi_t+\pi_s}\right\}\frac{1}{\pi_s}\\
	=& -\{\beta_t(z)- \beta_s(z)\}^{\top}  \Sigma_X \{\beta_t(z)- \beta_s(z)\} \left(\frac{1-\pi_t-\pi_s}{\pi_t+\pi_s}\right)\leq 0\nonumber
\end{align*}
This completes the proof for $ \sigma_A^2\leq \sigma_B^2 $, where the equality holds if and only if $\{\beta_t- \beta_s\}(1- \pi_t-\pi_s)=0$ and $ \sum_{t=1}^k \pi_t\beta_t= \{ \pi_s\beta_t+ \pi_t\beta_s\}/(\pi_t+\pi_s)  $.

\subsection{Proof of Theorem 2}

First, from $\bar X_t- \bar X= O_p(n^{-1/2})$ and $\hat \beta_t= \beta_t+o_p(1)$ from Lemma \ref{lemma: coef}, we have 
$
\hat\theta(\hat \beta_1, \dots, \hat \beta_k) = \hat\theta( \beta_1, \dots,  \beta_k)+o_p(n^{-1/2})$.  By using the definition $ \beta_t = \Sigma_X^{-1} {\rm cov} (X_i, Y_i^{(t)})$, we have 
\[
E\big[ X_i^{\top} \{ Y_i^{(t)}- \theta_t -(X_i- \mu_X)^{\top}  \beta_t \} \big]= {\rm cov} (X_i, Y_i^{(t)}) - {\rm cov} (X_i, Y_i^{(t)})=0.
\]
Because $ Z_i $ is discrete and $ X_i $ contains all joint levels of $Z_i$ as a sub-vector, according to the estimation equations from the  least squares, we have that 
\begin{eqnarray*}
	E\left[ I(Z_i=z)\{Y_i^{(t)} -\theta_t-(X_i- \mu_X)^{\top} \beta_t  \}\right]=0, ~ \forall z\in \mathcal{Z} ,
\end{eqnarray*}
and thus,
\begin{eqnarray}
	E\left\{Y_i^{(t)} -\theta_t-(X_i- \mu_X)^{\top} \beta_t \mid Z_i \right\}=0, ~ {\rm a.s.} \label{eq: conditional on Z} .
\end{eqnarray}
Moreover, recall that   $\mathcal{A}=\{A_1, \dots, A_n\}$ and  $\mathcal{F}=\{Z_1, \dots, Z_n\}$, then
\begin{align*}
	&E\left\{   \bar Y_t- \bar X_t^{\top} \beta_t  \mid  \mathcal{A}, \mathcal{F}  \right\}  = E\left\{ \frac{\sum_{i=1}^{n} I(A_i=a_t)  (Y_i^{(t)}- X_i^\top \beta_t)}{n_t} \mid  \mathcal{A}, \mathcal{F}  \right\}  \\
	& = \frac{\sum_{i=1}^{n} I(A_i=a_t) E\left\{   Y_i^{(t)}- X_i^\top \beta_t\mid Z_i \right\} }{n_t}  = \theta_t - \mu_X^\top \beta_t,  ~ {\rm a.s.}.
\end{align*}
This implies that $   \bar Y_t- \theta_t-( \bar X_t - \mu_X) ^{\top} \beta_t =  \bar Y_t- \bar X_t^{\top} \beta_t  - E( \bar Y_t- \bar X_t^{\top} \beta_t \mid\mathcal{A}, \mathcal{F}  ) $ a.s..

We decompose $\hat\theta( \beta_1, \dots,  \beta_k)$ as
\begin{align*}
	& \hat\theta( \beta_1, \dots,  \beta_k)- \theta= \left(
	\begin{array}{c}
		\bar Y_1- \theta_1- (\bar X_1- \mu_X)^{\top} \beta_1 \\
		\cdots\\
		\bar Y_k- \theta_k- (\bar X_k- \mu_X)^{\top} \beta_k
	\end{array}\right) +
	\left(
	\begin{array}{c}
		\beta_1^{\top}   (\bar X- \mu_X)  \\
		\cdots\\
		\beta_k^{\top}      (\bar X- \mu_X) 
	\end{array}\right) \\
	&=\underbrace{ \left(
		\begin{array}{c}
			\bar Y_1- \theta_1- (\bar X_1- \mu_X)^{\top} \beta_1 \\
			\cdots\\
			\bar Y_k- \theta_k- (\bar X_k- \mu_X)^{\top} \beta_k
		\end{array}\right)}_{M_1} +
	\underbrace{  \left(
		\begin{array}{c}
			\beta_1^{\top}   (\bar X-E(\bar X\mid \mathcal{A}, \mathcal{F} ))  \\
			\cdots\\
			\beta_k^{\top}      (\bar X-E(\bar X\mid \mathcal{A}, \mathcal{F} )) 
		\end{array}\right)}_{M_{21}} +
	\underbrace{   \left(
		\begin{array}{c}
			\beta_1^{\top}   (E(\bar X\mid \mathcal{A}, \mathcal{F} )- \mu_X)  \\
			\cdots\\
			\beta_k^{\top}      (E(\bar X\mid \mathcal{A}, \mathcal{F} )- \mu_X) 
		\end{array}\right) }_{M_{22}}
	\\
	& =   \left(
	\begin{array}{ccccc}
		a_1^{\top}  &  -	 \beta_1^{\top}  & {0}_p^{\top}  &\cdots & {0}_p^{\top}     \\
		\vdots &	\vdots & \vdots & \ddots  & \vdots\\
		a_k^{\top}  &  {0}_p^{\top}   &  {0}_p^{\top}  &\cdots & -  \beta_k^{\top} 
	\end{array}\right)
	\underbrace{  \left(
		\begin{array}{c}
			\bar Y_1-  E(\bar Y_1\mid \mathcal{A}, \mathcal{F} )\\
			\cdots\\
			\bar Y_k-  E(\bar Y_k\mid \mathcal{A}, \mathcal{F} )  \\
			\bar X_1-  E(\bar X_1\mid \mathcal{A}, \mathcal{F} )\\
			\cdots  \\
			\bar X_k-  E(\bar X_k\mid \mathcal{A}, \mathcal{F} )\\
		\end{array}\right)}_{\tilde V} \\
	&   +  
	\left(
	\begin{array}{ccc}
		\beta_1^{\top}&& \\
		&\ddots&\\
		&&\beta_k^{\top}
	\end{array}\right)
	\left(
	\begin{array}{ccccc}
		n^{-1} n_1I_p &  	n^{-1} n_2 I_p   &\dots & 	n^{-1} n_k  I_p \\
		\vdots &\vdots &\ddots &\vdots \\
		n^{-1} n_1I_p &  	n^{-1} n_2 I_p   &\dots & 	n^{-1} n_k  I_p  
	\end{array}\right) 
	\left(
	\begin{array}{c}
		\bar X_1-  E(\bar X_1\mid \mathcal{A}, \mathcal{F} )\\
		\vdots\\
		\bar X_k-  E(\bar Y_k\mid \mathcal{A}, \mathcal{F} ) 
	\end{array}\right)  + M_{22}. 
\end{align*}

Conditioned  on $ \mathcal{A}, \mathcal{F} $, every component in $ \tilde{V} $ is an average of independent terms. From Lindeberg's Central Limit Theorem, as $ n\rightarrow \infty $,  $ \sqrt{n} \tilde V  $ is asymptotically normal with mean 0 conditional on $  \mathcal{A}, \mathcal{F}  $, which implies  that $ \sqrt{n} (M_1+M_{21}) $   is asymptotically normal with mean 0 conditional  on $   \mathcal{A}, \mathcal{F} $.

Next, we calculate the variance. For $M_1$, the variance of  its $t$th component is  
\begin{eqnarray*}
	n {\rm var} (M_{1t}\mid \mathcal{A},\mathcal{F}) &=&  \frac{n}{n_t^2}  {\rm var} \left\{\sum_{i: A_i=a_t}Y_i^{(t)} -(X_i- \mu_X)^{\top} \beta_t \mid \mathcal{A},\mathcal{F}\right\} \\
	&=& \frac{n}{n_t^2} \sum_{i: A_i=a_t} {\rm var} \left\{Y_i^{(t)} -(X_i- \mu_X)^{\top} \beta_t \mid Z_i\right\} \\
	&=& \frac{n}{n_t^2} \sum_{z} \sum_{i: A_i=a_t, Z_i=z} {\rm var} \left\{Y_i^{(t)} -(X_i- \mu_X)^{\top} \beta_t \mid Z_i=z\right\} \\
	&=& \frac{n}{n_t} \sum_{z}\frac{ n_t(z)}{n_t} {\rm var} \left\{Y_i^{(t)} -(X_i- \mu_X)^{\top} \beta_t \mid Z_i=z\right\} \\
	&=& \frac{1}{\pi_t} \sum_{z} P(Z_i=z) {\rm var} \left\{Y_i^{(t)} -(X_i- \mu_X)^{\top} \beta_t \mid Z_i=z\right\} +o_p(1)\\
	&=& \frac{1}{\pi_t} E\left[ {\rm var} \left\{Y_i^{(t)} -(X_i- \mu_X)^{\top} \beta_t \mid Z_i\right\}\right] +o_p(1) ,
\end{eqnarray*}
where the second line  and the fifth line are respectively from (C2) (i) and (C2) (ii). Moreover, $ M_{1t} $ and $ M_{1s} $ are independent conditional on $ \mathcal{A}, \mathcal{F} $, for $ t\neq s $.  Hence, 
\begin{eqnarray}
	{\rm var} (	\sqrt{n} M_1\mid \mathcal{A},\mathcal{F} ) =  {\rm diag} \left\{  \pi_t^{-1} E\left[ {\rm var} \big\{Y_i^{(t)} -(X_i- \mu_X)^{\top} \beta_t \mid Z_i\big\}\right] \right\} + o_p(1) ,
\end{eqnarray}
which does not depend on the randomization scheme. For $ M_{21} $, we have that
\begin{align*}
	n {\rm var} ( \bar X - E(\bar X\mid \mathcal{A}, \mathcal{F})\mid \mathcal{A},\mathcal{F}) &= \frac{1}{n} \sum_{i=1}^{n} {\rm var} (X_i\mid Z_i) = E\{ {\rm var} (X_i\mid Z_i)  \} + o_p(1)\\
	n {\rm var} (M_{21}\mid \mathcal{A},\mathcal{F}) &=\mathscr{B}^\top  E\{ {\rm var} (X_i\mid Z_i)  \} \mathscr{B} + o_p(1) . 
\end{align*}
For the covariance, consider $n {\rm cov} (M_1,  M_{21}\mid \mathcal{A},\mathcal{F})$ whose $(t, s)$ element equals
\begin{eqnarray}
	&&n {\rm cov} (M_{1t},  \bar X^{\top} \beta_s\mid \mathcal{A},\mathcal{F})\label{eq: cov} \\
	&= &n {\rm cov} \left(\bar Y_t- \bar X_t ^\top \beta_t ,    \sum_{j=1}^{k} \frac{n_j}{n} \bar X_j^{\top} \beta_s\mid \mathcal{A},\mathcal{F}\right)  \nonumber \\
	&= &n {\rm cov} \left(\bar Y_t- \bar X_t ^\top \beta_t ,  \frac{n_t}{n} \bar X_t^{\top} \beta_s\mid \mathcal{A},\mathcal{F}\right)  \nonumber \\
	&= &\frac{1}{n_t} \sum_{i: A_i=a_t}{\rm cov} \left( Y_i^{(t) }-  X_i ^\top \beta_t ,    X_i^{\top} \beta_s\mid Z_i \right)  \nonumber \\
	&=& \frac{1}{n_t} \sum_{i: A_i=a_t} \sum_{z\in \mathcal{Z}} I(Z_i=z) {\rm cov} \left( Y_i^{(t) }-  X_i ^\top \beta_t ,    X_i^{\top} \beta_s\mid Z_i =z \right)  \nonumber \\
	&=& \sum_{z\in \mathcal{Z}}\frac{n_t(z)}{n_t} {\rm cov} \left( Y_i^{(t) }-  X_i ^\top \beta_t ,    X_i^{\top} \beta_s\mid Z_i =z \right)  \nonumber \\
	&=& \sum_{z\in \mathcal{Z}} P(Z=z) {\rm cov} \left( Y_i^{(t) }-  X_i ^\top \beta_t ,    X_i^{\top} \beta_s\mid Z_i =z \right)  + o_p(1) \nonumber \\
	&=& E\left\{  {\rm cov} \left( Y_i^{(t) }-  X_i ^\top \beta_t ,    X_i^{\top} \beta_s\mid Z_i \right)   \right\} + o_p(1) \nonumber \\
	&=& o_p(1) \nonumber, 
\end{eqnarray}
where the last  equality holds because $ E(Y_i^{(t)}- X_i^{\top}\beta_t \mid Z_i) = \theta_t- \mu_X^{\top} \beta_t$  and, thus,  $ {\rm cov}\{E(Y_i^{(t)}- X_i^{\top}\beta_t \mid Z_i) , E(X_i^{\top} \beta_s\mid Z_i) \} =0$ and $  E\{  {\rm cov} ( Y_i^{(t) }-  X_i ^\top \beta_t ,    X_i^{\top} \beta_s\mid Z_i )   \} =   {\rm cov} ( Y_i^{(t) }-  X_i ^\top \beta_t ,    X_i^{\top} \beta_s) =0 $ according to the definition of $ \beta_t $. 

Combining the above derivations and from the Slutsky's theorem, we have shown that 
\begin{align*}
	&\sqrt{n} (M_1+M_{21}) \mid\mathcal{A},\mathcal{F} \\
	& \xrightarrow{d} N\left(0,  {\rm diag} \left\{  \pi_t^{-1} E\left[ {\rm var} \big\{Y_i^{(t)} -(X_i- \mu_X)^{\top} \beta_t \mid Z_i\big\}\right] \right\}  +\mathscr{B}^\top  E\{ {\rm var} (X_i\mid Z_i)  \} \mathscr{B}   \right). 
\end{align*}
From the bounded convergence theorem, this result also holds unconditionally, i.e., 
\begin{align}
	&\sqrt{n} (M_1+M_{21})\nonumber\\
	& \xrightarrow{d} N\left(0,  {\rm diag} \left\{  \pi_t^{-1} E\left[ {\rm var} \big\{Y_i^{(t)} -(X_i- \mu_X)^{\top} \beta_t \mid Z_i\big\}\right] \right\}  +\mathscr{B}^\top  E\{ {\rm var} (X_i\mid Z_i)  \} \mathscr{B}   \right).  \nonumber
\end{align} 
Moreover, since $ E(\bar X\mid \mathcal{A}, \mathcal{F})  $ is an average of  identically and independently distributed  terms, by the central limit theorem, 
\begin{align}
	&\sqrt{n} \{ E(\bar X\mid \mathcal{A}, \mathcal{F}) - \mu_X \}= n^{-1/2} \sum_{i=1}^{n}   \{ E( X_i\mid Z_i) - \mu_X \}  \xrightarrow{d} N(0, \var (E(X_i\mid Z_i)) ), \nonumber
\end{align}
and 
\begin{align*}
	&\sqrt{n} M_{22}  \xrightarrow{d} N(0,  \mathscr{B}^\top \var (E(X_i\mid Z_i)) \mathscr{B} ).\nonumber
\end{align*}

Next, we show that $ (\sqrt{n} (M_1+M_{21})  , \sqrt{n} M_{22}   ) \xrightarrow{d} (\xi_1, \xi_2)$, where $ (\xi_1,\xi_2) $ are mutually independent. This can be seen from 
\begin{align*}
	&P(\sqrt{n} (M_1+M_{21}) \leq t_1 , \sqrt{n} M_{22}  \leq t_2)  \\
	&= E\{ I(\sqrt{n} (M_1+M_{21}) \leq t_1) I( \sqrt{n} M_{22}  \leq t_2) \} \\
	&= E\{ P(\sqrt{n} (M_1+M_{21}) \leq t_1\mid \mathcal{A}, \mathcal{F}) I( \sqrt{n} M_{22}  \leq t_2) \} \\
	&= E\{ \{ P(\sqrt{n} (M_1+M_{21}) \leq t_1\mid \mathcal{A}, \mathcal{F}) - P(\xi_1\leq t_1) \}  I( \sqrt{n} M_{22}  \leq t_2) \}  \\
	&\qquad +P(\xi_1\leq t_1) P(\sqrt{n} M_{22}\leq t_2)  \\
	&\rightarrow P(\xi_1\leq t_1) P(\xi_2 \leq t_2)  ,
\end{align*}
where the last step follows from the bounded convergence theorem. 

Finally, from $ \sqrt{n} \{  \hat\theta( \beta_1, \dots,  \beta_k)- \theta \}  = \sqrt{n} (M_1+M_{21}+M_{22}) $, we have 
\begin{align*}
	&	\sqrt{n} \{  \hat\theta( \beta_1, \dots,  \beta_k)- \theta \} \\
	& \xrightarrow{d} N\left(0, {\rm diag} \left\{  \pi_t^{-1} E\left[ {\rm var} \big\{Y_i^{(t)} -(X_i- \mu_X)^{\top} \beta_t \mid Z_i\big\}\right] \right\}  +  \mathscr{B}^{\top}\Sigma_X  \mathscr{B}\right). 
\end{align*}

Note that we have also shown that the asymptotic distribution of $  \sqrt{n}\{  \hat\theta( \beta_1, \dots,  \beta_k)- \theta\}$ is invariant under randomization schemes satisfying (C2).   The above asymptotic distribution of $  \sqrt{n}\{  \hat\theta( \beta_1, \dots,  \beta_k)- \theta\}$ is  the same as \eqref{normal1} because $ E\{ Y_i^{(t)} - \theta_t- (X_i- \mu_X)^{\top} \beta_t\mid Z_i\}= 0$ a.s.,  and thus, $ E[ {\rm var} \{Y_i^{(t)} -(X_i- \mu_X)^{\top} \beta_t \mid Z_i\}]  = {\rm  var} (Y_i^{(t)}- X_i^{\top} \beta_t)$.

\subsection{Proof of Theorem 3}
(i) First, from $\bar X_t- \bar X= O_p(n^{-1/2})$ and $\hat b_t= b_t+o_p(1)$, we have 
$
\hat\theta(\hat b_1, \dots, \hat b_k) = \hat\theta( b_1, \dots,  b_k)+o_p(n^{-1/2})$.  Also note that 
\begin{align*}
	&E(\bar Y_t- \theta_t- (\bar X_t - \mu_X)^\top b_t \mid \mathcal{A}, \mathcal{F}) \\
	&= E\left(\frac{\sum_{i=1}^{n} I(A_i=a_t) (Y_i^{(t)} - \theta_t -( X_i- \mu_X)^\top b_t) }{n_t} \mid \mathcal{A}, \mathcal{F} \right) \\
	&=  \frac{\sum_{i=1}^{n} (I(A_i=a_t) - \pi_t )E( Y_i^{(t)} - \theta_t -( X_i- \mu_X)^\top b_t\mid Z_i)  }{n_t} \\
	&\qquad + \frac{\pi_t}{n_t} \sum_{i=1}^n E(Y_i^{(t)} -\theta_t - (X_i- \mu_X) ^\top b_t  \mid Z_i) \\
	&= \sum_{z\in \mathcal{Z}} \left( \frac{n_t(z)}{n(z)  } - \pi_t \right) E(Y_i^{(t)} -\theta_t- (X_i- \mu_X)^\top b_t \mid Z_i =z ) \frac{n(z)}{n_t} \\
	&\qquad + \frac{\pi_t}{n_t} \sum_{i=1}^n E(Y_i^{(t)} -\theta_t- (X_i- \mu_X)^\top b_t  \mid Z_i) \\
	&=  \sum_{z\in \mathcal{Z}} \left( \frac{n_t(z)}{n(z)  } - \pi_t \right) E(Y_i^{(t)}  - \theta_t- (X_i- \mu_X)^\top b_t \mid Z_i =z ) P(Z=z)\pi_t^{-1} \\
	&\qquad +  \frac{1}{n} \sum_{i=1}^n E(Y_i^{(t)} -\theta_t- (X_i- \mu_X)^\top b_t  \mid Z_i)  +o_p(n^{-1/2}), 
\end{align*}
where the last equality is from  $ n(z)/n= P(Z=z)+o_p(1) $, $ n_t/n= \pi_t+o_p(1) $,  $  \left( \frac{n_t(z)}{n(z)  } - \pi_t \right)  =O_p(n^{-1/2})$ due to condition (C3), and $n^{-1} \sum_{i=1}^n E(Y_i^{(t)} -\theta_t - (X_i- \mu_X) ^\top b_t  \mid Z_i) = O_p(n^{-1/2}) $. 

Thus, we can decompose $\hat\theta( b_1, \dots,  b_k)$ as
\begin{align*}
	& \hat\theta( b_1, \dots,  b_k)- \theta\\
	&= \left(
	\begin{array}{c}
		\bar Y_1- \theta_1- (\bar X_1- \mu_X)^{\top} b_1 \\
		\cdots\\
		\bar Y_k- \theta_k- (\bar X_k- \mu_X)^{\top} b_k
	\end{array}\right) +
	\left(
	\begin{array}{c}
		b_1^{\top}   (\bar X- \mu_X)  \\
		\cdots\\
		b_k^{\top}      (\bar X- \mu_X) 
	\end{array}\right) \\
	&=\underbrace{ \left(
		\begin{array}{c}
			\bar Y_1-  E(\bar Y_1\mid \mathcal{A}, \mathcal{F} ) - (\bar X_1- E(\bar X_1\mid \mathcal{A}, \mathcal{F} ) )^{\top} b_1 \\
			\cdots\\
			\bar Y_k-  E(\bar Y_k\mid \mathcal{A}, \mathcal{F} ) - (\bar X_k- E(\bar X_k\mid \mathcal{A}, \mathcal{F} ) )^{\top} b_k
		\end{array}\right)}_{M_{11}} +
	\underbrace{  \left(
		\begin{array}{c}
			b_1^{\top}   (\bar X-E(\bar X\mid \mathcal{A}, \mathcal{F} ))  \\
			\cdots\\
			b_k^{\top}      (\bar X-E(\bar X\mid \mathcal{A}, \mathcal{F} )) 
		\end{array}\right)}_{M_{21}}  \\
	&+ 
	\underbrace{ \left(
		\begin{array}{c}
			\sum_{z\in \mathcal{Z}} \left( \frac{n_1(z)}{n(z)  } - \pi_1 \right) E(Y_i^{(1)} -\theta_1- (X_i- \mu_X)^\top b_1 \mid Z_i =z ) P(Z=z)\pi_1^{-1} \\
			\cdots\\
			\sum_{z\in \mathcal{Z}} \left( \frac{n_k(z)}{n(z)  } - \pi_k \right) E(Y_i^{(k)} -\theta_k- (X_i-\mu_X)^\top b_k \mid Z_i =z ) P(Z=z)\pi_k^{-1} 
		\end{array}\right)}_{M_{12}}
	\\
	&+ 
	\underbrace{ \left(
		\begin{array}{c}
			n^{-1} \sum_{i=1}^n E(Y_i^{(1)} -\theta_1- (X_i - \mu_X)^\top b_1 \mid Z_i) \\
			\cdots\\
			n^{-1}  \sum_{i=1}^n E(Y_i^{(k)} -  \theta_k  - (X_i - \mu_X)^\top b_k  \mid Z_i)
		\end{array}\right)}_{M_{31}} + 
	\underbrace{ \left(
		\begin{array}{c}
			n^{-1} \sum_{i=1}^n b_1^\top E(X_i-\mu_X \mid Z_i) \\
			\cdots\\
			n^{-1}  \sum_{i=1}^n b_k^\top E(X_i-\mu_X \mid Z_i)  
		\end{array}\right)}_{M_{32}} \\
	& +o_p(n^{-1/2})
	\\
	& =   \left(
	\begin{array}{ccccc}
		a_1^{\top}  &  -	 b_1^{\top}  & {0}_p^{\top}  &\cdots & {0}_p^{\top}     \\
		\vdots &	\vdots & \vdots & \ddots  & \vdots\\
		a_k^{\top}  &  {0}_p^{\top}   &  {0}_p^{\top}  &\cdots & -  b_k^{\top} 
	\end{array}\right)
	\underbrace{  \left(
		\begin{array}{c}
			\bar Y_1-  E(\bar Y_1\mid \mathcal{A}, \mathcal{F} )\\
			\cdots\\
			\bar Y_k-  E(\bar Y_k\mid \mathcal{A}, \mathcal{F} )  \\
			\bar X_1-  E(\bar X_1\mid \mathcal{A}, \mathcal{F} )\\
			\cdots  \\
			\bar X_k-  E(\bar X_k\mid \mathcal{A}, \mathcal{F} )\\
		\end{array}\right)}_{\tilde V_1 } \\
	&   +  
	\left(
	\begin{array}{ccc}
		b_1^{\top}&& \\
		&\ddots&\\
		&&b_k^{\top}
	\end{array}\right)
	\left(
	\begin{array}{ccccc}
		n^{-1} n_1I_p &  	n^{-1} n_2 I_p   &\dots & 	n^{-1} n_k  I_p \\
		\vdots &\vdots &\ddots &\vdots \\
		n^{-1} n_1I_p &  	n^{-1} n_2 I_p   &\dots & 	n^{-1} n_k  I_p  
	\end{array}\right) 
	\left(
	\begin{array}{c}
		\bar X_1-  E(\bar X_1\mid \mathcal{A}, \mathcal{F} )\\
		\vdots\\
		\bar X_k-  E(\bar Y_k\mid \mathcal{A}, \mathcal{F} ) 
	\end{array}\right)  \\
	&+ M_{12} + M_{31} +M_{32} +o_p(n^{-1/2}) .
\end{align*}

Conditioned  on $ \mathcal{A}, \mathcal{F} $, every component in $ \tilde{V}_1 $ is an average of independent terms. From the Lindeberg's Central Limit Theorem, as $ n\rightarrow \infty $,  $ \sqrt{n} \tilde V_1  $ is asymptotically normal with mean 0 conditional on $  \mathcal{A}, \mathcal{F}  $, which combined with the Cram\'{e}r-Wold device  implies  that $ \sqrt{n} (M_{11}+M_{21}) $   is asymptotically normal with mean 0 conditional  on $   \mathcal{A}, \mathcal{F} $. Following the same steps as in the proof of Theorem 2, we have that 
\begin{align}
	&\sqrt{n} ( M_{11}+M_{21}) \mid \mathcal{A}, \mathcal{F} \xrightarrow{d} \nonumber \\
	&  N\bigg(0,  {\rm diag} \left\{  \pi_t^{-1} E[ {\rm var} \{Y_i^{(t)} -X_i^{\top} b_t \mid Z_i\}] \right\}  +{B}^\top  E\{ {\rm var} (X_i\mid Z_i)  \} {B} \nonumber\\
	& \qquad\qquad + (\mathscr{B}  - B)^\top  E\{ {\rm var} (X_i\mid Z_i)  \} B+ {B}^\top  E\{ {\rm var} (X_i\mid Z_i)  \} (\mathscr{B}  - B) \bigg),  \nonumber 
\end{align}   
and
\begin{align}
	&\sqrt{n} ( M_{11}+M_{21})  \xrightarrow{d} \nonumber\\
	&  N\bigg(0,  {\rm diag} \left\{  \pi_t^{-1} E[ {\rm var} \{Y_i^{(t)} -X_i^{\top} b_t \mid Z_i\}] \right\}  +{B}^\top  E\{ {\rm var} (X_i\mid Z_i)  \} {B} \nonumber\\
	& \qquad\qquad + (\mathscr{B}  - B)^\top  E\{ {\rm var} (X_i\mid Z_i)  \} B+ {B}^\top  E\{ {\rm var} (X_i\mid Z_i)  \} (\mathscr{B}  - B) \bigg).  \nonumber 
\end{align}

Next,  notice that $ \sqrt{n}M_{12} $ is asymptotically normal  conditional on $ \mathcal{F} $ with mean 0 from condition (C3). Let  $ \omega_{ts}(z) $ be the $ (t, s) $ element in the matrix $ \Omega(z) $,  then the conditional variance of  $\sqrt{n}  M_{12t} $,  the $ t $th component of $\sqrt{n}  M_{12} $, equals 
\begin{eqnarray*}
	&&{\rm var}(    \sqrt{n} M_{12t}  \mid \mathcal{F})\\
	&=&\pi_t^{-2}\sum_{z} \bigg[E\left\{Y_i^{(t)}- \theta_t -(X_i- \mu_X)^{\top} b_t\mid Z=z\right\}\bigg]^2 P(Z_i=z)  {\rm var} \left\{ \frac{n_t(z)- \pi_t n(z) }{\sqrt{n(z)}} \mid \mathcal{F} \right\} \\
	&&+ o_p(1)\\
	&=&\pi_t^{-2}\sum_{z} \bigg[E\left\{Y_i^{(t)}- \theta_t -(X_i- \mu_X)^{\top} b_t\mid Z_i=z\right\}\bigg]^2 P(Z_i=z) \omega_{tt}(z)+ o_p(1)\\
	&=& \pi_t^{-2} E \left[ \omega_{tt}(Z) \big[ E\{Y_i^{(t)}- \theta_t -(X_i- \mu_X)^{\top} b_t\mid Z_i\} \big]^2\right]+o_p(1) ,
\end{eqnarray*}
and the conditional covariance between $ \sqrt{n}  M_{12t}   $ and  $ \sqrt{n}  M_{12s}   $ equals 
\begin{eqnarray*}
	&&      {\rm cov}(      \sqrt{n} M_{12t} , \sqrt{n} M_{12s}\mid \mathcal{F}) \\
	&&=  \frac{1}{\pi_t\pi_s} \sum_z \prod_{m \in \{t, s\} } E\left\{Y_i^{(m)}- \theta_i -(X_i- \mu_X)^{\top} b_m\mid Z=z\right\} P(Z_i=z) \\
	&&\qquad\qquad  {\rm cov} \left\{ \frac{n_t(z)- \pi_t n(z) }{\sqrt{n(z)}} , \frac{n_s(z)- \pi_s n(z) }{\sqrt{n(z)}} \mid \mathcal{F}\right\} +o_p(1) \\
	&&= \frac{1}{\pi_t\pi_s}  E \left[ \omega_{ts}(Z) E\left\{Y_i^{(t)}- \theta_t -(X_i- \mu_X)^{\top} b_t\mid Z_i\right\}E\left\{Y_i^{(s)}- \theta_s -(X_i- \mu_X)^{\top} b_s\mid Z_i\right\} \right] \\
	&& \qquad \qquad+o_p(1) .
\end{eqnarray*}
Therefore, from the Slutsky's theorem, 
\begin{align}
	\sqrt{n}M_{12} \mid \mathcal{F} \xrightarrow{d} N\big(0, E\left\{ R(B) \Omega(Z_i) R(B) \right\} \big) \nonumber.
\end{align}

Moreover,  $M_{31} +M_{32}$ only involves sums of identically and independently distributed terms, and $E(M_{31} +M_{32})=0$. Again using the  Cram\'{e}r-Wold device similarly to the proof of $ M_{11}+M_{21} $, we have that $\sqrt{n} (M_{31} +M_{32})  $ is asymptotically normal. Let $ \pi= (\pi_1, \dots, \pi_k)^{\top} $, it is easy to show  that 
\begin{eqnarray*}
	\var (\sqrt{n} M_{31} )   = {\rm var} \{ R(B) \pi \} =  E\{ R(B) \pi \pi^\top R(B)\} , \quad \var (\sqrt{n} M_{32} )   = B^\top  \var\{ E(X_i  \mid Z_i)\}  B, 
\end{eqnarray*}
and the $ (t,s) $ component of ${\rm cov} (\sqrt{n} M_{31}, \sqrt{n}M_{32}) $ is 
\begin{align*}
	{\rm cov} (\sqrt{n} M_{31t}, \sqrt{n}M_{32s}) &=  {\rm cov}\left[ E \left\{Y_i^{(t)} -(X_i- \mu_X)^{\top} b_t\mid Z_i \right\} , E( X_i^{\top} b_s\mid Z_i)  \right]  \\
	&= (\beta_t- b_t)^\top \var (E(X_i \mid Z_i ))  b_s .
\end{align*}
Hence,
\begin{align*}
	& \var (\sqrt{n} (M_{31} + M_{32}))  \\
	&=  E\{ R(B) \pi \pi^\top R(B)\} + B^\top  \var\{ E(X_i  \mid Z_i)\}  B+ (\mathscr{B}- B)^{\top}  \var (E(X_i \mid Z_i ))   B  \\
	&\qquad + B^\top \var (E(X_i \mid Z_i ))   (\mathscr{B}- B) . 
\end{align*}
Therefore, 
\begin{align}
	&	\sqrt{n} (M_{31} + M_{32})  \label{eq: M_3}\\
	&\xrightarrow{d} N\bigg(0,  E\{ R(B) \pi \pi^\top R(B)\} + B^\top  \var\{ E(X_i  \mid Z_i)\}  B+ (\mathscr{B}- B)^{\top}  \var (E(X_i \mid Z_i ))   B  \nonumber \\
	&  \qquad \qquad  + B^\top \var (E(X_i \mid Z_i ))   (\mathscr{B}- B)  \bigg).  \nonumber
\end{align}

Next, we show that $ (\sqrt{n}(M_{11} + M_{21}), \sqrt{n}M_{12}, \sqrt{n}(M_{31}+M_{32}))\xrightarrow{d} (\xi_{M1}, \xi_{M2}, \xi_{M3}) $, where $ (\xi_{M1}, \xi_{M2}, \xi_{M3}) $ are mutually independent. This can be seen from 
\begin{align*}
	&P(\sqrt{n}(M_{11} + M_{21})\leq t_1, \sqrt{n}M_{12}\leq t_2, \sqrt{n}  (M_{31}+M_{32})\leq t_3) \\
	=&  E\left\{ I(\sqrt{n}(M_{11} + M_{21})\leq t_1) I( \sqrt{n}M_{12}\leq t_2) I(\sqrt{n} (M_{31}+M_{32})\leq t_3) \right\}\\
	=&  E\left\{ P(\sqrt{n}(M_{11} + M_{21})\leq t_1\mid \mathcal{A}, \mathcal{F}) I( \sqrt{n}M_{12}\leq t_2) I(\sqrt{n}  (M_{31}+M_{32}) \leq t_3) \right\}\\
	=& E\left[ \big\{P(\sqrt{n}(M_{11} + M_{21})\leq t_1\mid \mathcal{A}, \mathcal{F})- P(\xi_{M1} \leq t_1) \big\} I( \sqrt{n} M_{12}\leq t_2) I(\sqrt{n}  (M_{31}+M_{32}) \leq t_3) \right]\\
	& \qquad + P(\xi_{M1} \leq t_1) E\left\{ I( \sqrt{n}M_{12}\leq t_2) I(\sqrt{n}  (M_{31}+M_{32})  \leq t_3) \right\}\\
	=&  E\left[ \big\{P(\sqrt{n}(M_{11} + M_{21}) \leq t_1\mid \mathcal{A}, \mathcal{F})- P(\xi_{M1} \leq t_1) \big\} I( \sqrt{n}M_{12}\leq t_2) I(\sqrt{n}  (M_{31}+M_{32}) \leq t_3) \right]\\
	& \qquad + P(\xi_{M1} \leq t_1) E\left\{ \left[P( \sqrt{n}M_{12}\leq t_2 \mid \mathcal{F} )- P(\xi_{M2} \leq t_2)\right]I(\sqrt{n}  (M_{31}+M_{32}) \leq t_3) \right\} \\
	&\qquad + P(\xi_{M1} \leq t_1)  P(\xi_{M2} \leq t_2) P(\sqrt{n}  (M_{31}+M_{32})  \leq t_3) \\
	\rightarrow & P(\xi_{M1} \leq t_1)  P(\xi_{M2} \leq t_2) P(\xi_{M3}\leq t_3), 
\end{align*}
where the last step follows from the bounded convergence theorem.

Finally, from $ \sqrt{n} \{  \hat\theta( b_1, \dots,  b_k)- \theta \}  = \sqrt{n} (M_{11}+M_{21}+M_{12}+ M_{31}+ M_{32}) +o_p(n^{-1/2})$, we conclude that  $ \sqrt{n}(\hat\theta (b_1, \dots, b_k)  - \theta )$ is asymptotically normal with mean 0 and variance
\begin{eqnarray}
	&& {\rm diag} \left\{  \pi_t^{-1} E\{ {\rm var} (Y_i^{(t)} -X_i^{\top} b_t \mid Z_i)\} \right\}+ E\left\{ R(B) \Omega(Z_i) R(B) \right\} + E\left\{R(B) \pi \pi^{\top} R(B) \right\}  \nonumber \\
	&&+ B^{\top} \Sigma_X B +  (\mathscr{B}- B)^{\top} \Sigma_X B + B^{\top} \Sigma_X  (\mathscr{B}- B) \nonumber\\
	&=& {\rm diag} \left\{  \pi_t^{-1} E\{ {\rm var} (Y_i^{(t)} -X_i^{\top} b_t \mid Z_i)\} \right\}+ E\left\{ R(B) \Omega(Z_i) R(B) \right\} + E\left\{R(B) \pi \pi^{\top} R(B) \right\}  \nonumber \\
	&&-  B^{\top} \Sigma_X B +  \mathscr{B}^{\top} \Sigma_X B + B^{\top} \Sigma_X  \mathscr{B}\nonumber\\
	&=& {\rm diag} \left\{ \pi_t^{-1} {\rm var}  ( Y^{(t)} - b_t^{\top} X) \right\}-  {\rm diag} \left\{  \pi_t^{-1}{\rm var} \{E (Y_i^{(t)} -X_i^{\top} b_t \mid Z_i)\} \right\}\nonumber\\
	&&+ E\left\{ R(B) \Omega(Z_i) R(B) \right\} + E\left\{R(B) \pi \pi^{\top} R(B) \right\} -  B^{\top} \Sigma_X B +  \mathscr{B}^{\top} \Sigma_X B + B^{\top} \Sigma_X  \mathscr{B}\nonumber\\
	&=& V_{\rm SR}  (B) - {\rm diag} \left\{  \pi_t^{-1}{\rm var} \{E (Y_i^{(t)} -X_i^{\top} b_t \mid Z_i)\} \right\} +  E\left[ R(B) \{\Omega(Z_i) + \pi\pi^{\top}\} R(B) \right] \nonumber \\
	&=&  V_{\rm SR}  (B) - {\rm diag} \left\{  R(B) {\rm diag} ( \pi_t) R(B)\right\} +  E\left[ R(B) \{\Omega(Z_i) + \pi\pi^{\top}\} R(B) \right] \nonumber    \\
	&=&  V_{\rm SR}  (B)  -  E\left[ R(B) \{{\rm diag} ( \pi_t) -  \pi\pi^{\top} - \Omega(Z_i)\} R(B) \right] \nonumber   \\
	&=&  V_{\rm SR}  (B)  -  E\left[ R(B) \{\Omega_{\rm SR}- \Omega(Z_i)\} R(B) \right] \nonumber   . 
\end{eqnarray}
(ii)
By using the definition $\beta_t =
\Sigma_X^{-1} {\rm cov}(X_i,Y_i^{(t)})$, we have 
$ 	E[X_i^{\top} \{Y_i^{(t)} -
\theta_t- \beta_t^{\top}(X- \mu_X)\} ]
= {\rm
	cov}(X_i,Y_i^{(t)}) - {\rm cov}(X_i,Y_i^{(t)})
= 0. $
Because $X_i$ contains all dummy variables for
the joint levels of $Z_i$,
we have $E\{Y_i^{(t)}-  \theta_t-  \beta_t^{\top} (X_i- \mu_X)\mid Z_i\}= 0$. Hence
$R(\mathscr{B}) = 0$ and $R(B)= {\rm diag} \{\pi_t^{-1} (\beta_t - b_t)^{\top} E(X_i - \mu_X \mid Z_i)\}$. Consequently,  the difference in asymptotic
variance is
\[
\begin{split}
	&V(B)- V(\mathscr{B})=V(B)-  V_{\rm SR}(\mathscr{B})=V(B)- V_{\rm SR}(B)+  V_{\rm SR}(B)-
	V_{\rm SR}(\mathscr{B})\\
	=& \text{diag}\{\pi_t^{-1}
	(\beta_t - b_t)^{\top} \Sigma_X(\beta_t - b_t) \} -  (\mathscr{B}-
	B)^{\top}\Sigma_X(\mathscr{B}- B) -  E\left[ R(B) \{  \Omega_{\rm SR}-
	\Omega (Z_i) \} R(B)\}\right] \\
	\ge & \text{diag}\{\pi_t^{-1}
	(\beta_t - b_t)^{\top} \Sigma_X(\beta_t - b_t) \} -  (\mathscr{B}-
	B)^{\top}\Sigma_X(\mathscr{B}- B) -  E\left[ R(B)  \Omega_{\rm SR}
	R(B)\}\right]\\
	=&  {\rm diag}
	[\pi_t^{-1}  (\beta_t- b_t)^{\top}  E \{{\rm var}(X\mid Z)\} (\beta_t-
	b_t)]- (\mathscr{B}- B)^{\top}E \{{\rm var}(X\mid Z)\} (\mathscr{B}- B),
\end{split}
\]
where $M \ge M'$ means $M - M'$ is positive semidefinite for two
square matrices $M$ and $M'$ of the same dimension, and the last line follows from
$\Omega_{\rm SR} = {\rm diag} (\pi_t)- \pi \pi^{\top}$, the
expression for $R(B)$, and the
identity $\Sigma_X = E\{{\rm var}(X \mid Z)\} + {\rm var}\{E(X \mid
Z)\}$. 	The positive semidefiniteness of the right hand side is from applying Lemma 1 with $ M= [E \{{\rm var}(X\mid Z)\} ]^{1/2} (\mathscr{B}- B) $.

\subsection{Proof of Corollary 2} 
When $ X $ \emph{only} contains the dummy variables for the joint levels of $ Z $, $ R(B)={\rm diag} \{\pi_t^{-1} (\beta_t - b_t)^{\top}( X_i - \mu_X )\} $. Then, it follows from the proof of Theorem 3(ii) that 
\begin{align*}
	& 	V(B)- V_{\rm SR}(\mathscr{B}) \\
	=& \text{diag}\{\pi_t^{-1} (\beta_t - b_t)^{\top} \Sigma_X(\beta_t - b_t) \} -  (\mathscr{B}-
	B)^{\top}\Sigma_X(\mathscr{B}- B) -  E\left[ R(B) \Omega_{\rm SR}R(B)\}\right] \\
	&\qquad  + E\left[ R(B) \Omega(Z_i)R(B)\}\right]  \\
	=& \text{diag}\{\pi_t^{-1} (\beta_t - b_t)^{\top} \Sigma_X(\beta_t - b_t) \} -  (\mathscr{B}-
	B)^{\top}\Sigma_X(\mathscr{B}- B) -  E\left[ R(B) {\rm diag}(\pi_t)R(B)\}\right] \\
	&\qquad  +   E\left[ R(B) \pi \pi^{\top}R(B)\}\right]+ E\left[ R(B) \Omega(Z_i)R(B)\}\right]  \\
	=& E[ R(B) \Omega(Z_i) R(B)]
\end{align*}

\end{document}